\documentclass{aa}

\usepackage{graphicx}
\usepackage{subcaption}
\usepackage{txfonts}
\begin{document}

   \title{Protostellar disk accretion in turbulent filaments}

   \subtitle{}

   \author{S. Heigl\inst{1,2}
          \and
          E. Hoemann\inst{2,3}
          \and
          A. Burkert\inst{1,2,3}
          }

   \institute{Universit\"ats-Sternwarte, Ludwig-Maximilians-Universit\"at M\"unchen,
             Scheinerstr. 1, 81679 Munich, Germany\\
             \email{heigl@usm.lmu.de}
         \and
             Excellence Cluster ORIGINS,
             Boltzmannstrasse 2, 85748 Garching, Germany
         \and
             Max-Planck Institute for Extraterrestrial Physics,
             Giessenbachstr. 1, 85748 Garching, Germany
             }

   \date{Received; accepted}

% \abstract{}{}{}{}{}
% 5 {} token are mandatory

  \abstract
   {Recent observations of protostellar cores suggest that most of the material
   in the protostellar phase is accreted along streamers. Streamers in this context
   are defined as velocity coherent funnels of denser material potentially connecting
   the large scale environment to the small scales of the forming accretion disk.}
   {Using simulations which simultaneously resolve the driving of turbulence on the
   filament scale as well as the collapse of the core down to protostellar disk scales,
   we aim to understand the effect of the turbulent velocity field on the formation of
   overdensities in the accretion flow.}
   {We perform a three-dimensional numerical study on a core collapse within a turbulent
   filament using the RAMSES code and analyse the properties of overdensities in the
   accretion flow.}
   {We find that overdensities are formed naturally by the initial turbulent velocity
   field inherited from the filament and subsequent gravitational collimation. This
   leads to streams which are not really filamentary but show a sheet-like morphology.
   Moreover, they have the same radial infall velocities as the low density material.
   As a main consequence of the turbulent initial condition, the mass accretion onto
   the disk does not follow the predictions for solid body rotation. Instead, most of
   the mass is funneled by the overdensities to intermediate disk radii.}
   {}


   \maketitle
%
%________________________________________________________________

\section{Introduction}

   While the classical picture of star formation assumes the gravitational collapse of an isolated
   spherical stellar core similar to a Bonnor-Ebert or singular isothermal sphere
   \citep{larson1969, penston1969, shu1977, foster1993}, it is now well supported that most star forming
   cores lie within filaments of molecular clouds \citep{andre2014}. Consequently, stars do not form
   in isolation but can accrete actively from their environment. This picture also has been established
   by observations with the emergence of streamers, velocity coherent non-isotropic accretion
   structures, which have been detected during all stages in the evolution of young stellar objects
   \citep{pineda2023}.

   During the disk build-up in embedded protostellar class-0/I objects, non-isotropic accretion
   streams have been observed in dust \citep{legouellec2019} and molecular emission \citep{cabedo2021,
   murillo2022, valdiviamena2022, kido2023}. Typically, the measured sizes are around a few thousand
   AU with accretion rates of around $1.0 \times 10^{-6}\,\mathrm{M_\sun\,yr^{-1}}$.
   However, there have been cases of streamers extending up to $10000\,\mathrm{AU}$ \citep{pineda2020}
   and accretion rates exceeding $1.0 \times 10^{-5}\,\mathrm{M_\sun\,yr^{-1}}$ \citep{valdiviamena2023}.
   Most observations show an one major accretion stream onto a single core.
   Nonetheless, there have been detections of more complex structures such as multiple streamers
   \citep{thieme2022} and streamers accreting onto binary system \citep{hsieh2023}.

   Non-isotropic infall has a major impact on the forming protoplanetary disk, potentially leading to
   various substructures and instabilities \citep{bae2015, lesur2015, kuffmeier2021, kuznetsova2022}.
   This raises the fundamental question if planet formation already begins during the build-up of the
   disk in the class-0 phase in contrast to an isolated unperturbed disk in the class-II phase which
   is often assumed \citep{testi2014, lesur2022}. Not only was it shown by \citet{drazkowska2018} that
   planetesimals may form during infall, but there is even growing observational evidence that planet
   formation may start at an earlier stage. Submillimeter observations of disks in the class-II phase
   show insufficient dust masses to explain the formation of exoplanetary systems \citep{najita2014,
   ansdell2017, manara2018} in contrast to dust masses of disks in the class-0/I phase
   \citep{greaves2011, tychoniec2018, tobin2020, tychoniec2020}. Additionally, there are indications of
   dust growth in the early phases of disk formation \citep{kwon2009, harsono2018, hsieh2019,
   seguracox2020}. Finally, the isotopic distribution in meteorites \citep{kruijer2014, vankooten2016,
   kruijer2017} and lack of water in the inner Solar System \citep{morbidelli2016} suggest the early
   formation of a dynamical barrier such as Jupiter.

   \begin{figure*}
     \centering
     \includegraphics[width=1.8\columnwidth]{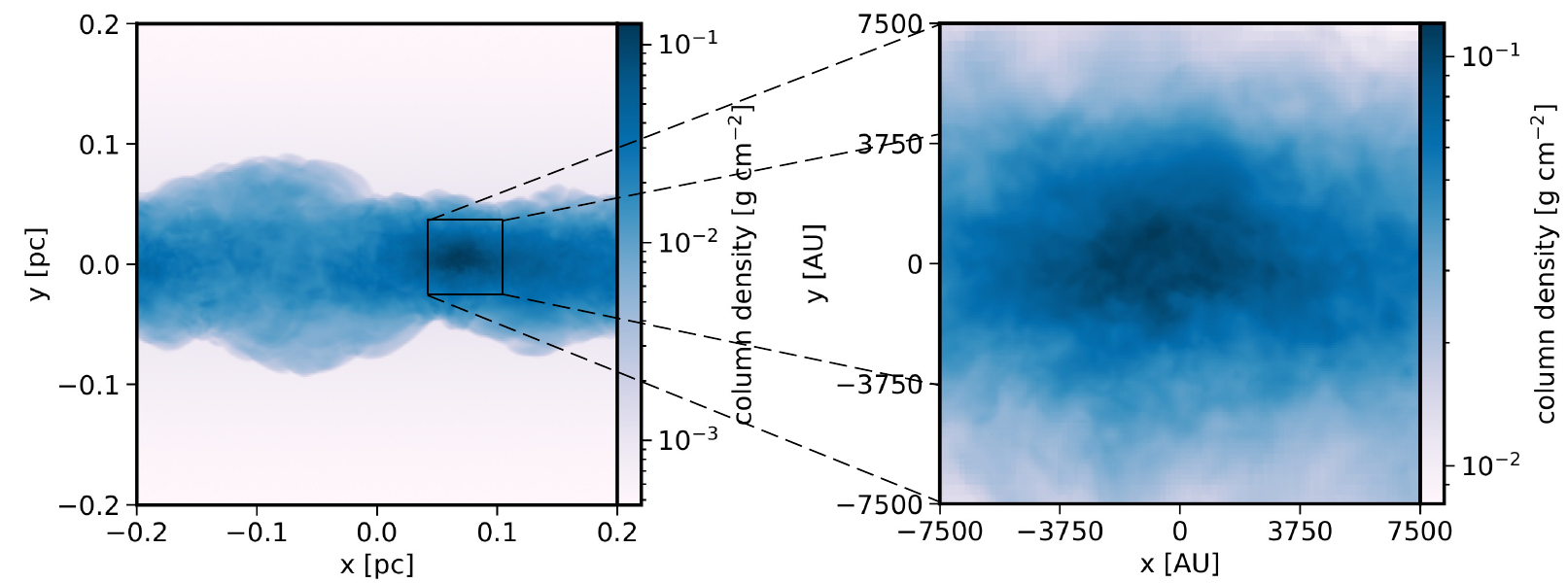}
     \caption{Projection of the core at $93.2\,\mathrm{kyr}$ before its
     collapse and sink particle creation. The left hand side shows the
     projection of the whole box with a size of $0.4\,\mathrm{pc}$. The
     right hand side shows a zoom onto the core with a radius of
     $7500\,\mathrm{AU}$. One can see that the core shows substructure
     as it inherits turbulent motions from the large scale filament.}
     \label{fig:core}
   \end{figure*}

   Several numerical studies have also shown that the large scales play an essential role for the
   distribution of mass accretion onto the disk. Models of collapsing cores including turbulence and
   magnetic fields usually lead to the formation of accretion channels onto the disk \citep{walch2010,
   joos2013, seifried2013, seifried2015, matsumoto2017, lam2019, hennebelle2020}. These can have a major
   impact on the accretion source function which specifies where accreted material enters the disk. For
   example, \citet{lee2021} detected larger accretion rates towards the centre than expected from
   angular momentum conservation of solid body rotation. Some studies were also able to
   self-consistently resolve the dynamical range from molecular clouds down to the
   disk using either adaptive mesh refinement \citep{kuffmeier2017, kuffmeier2018, kuffmeier2019,
   lebreuilly2021, pelkonen2021, kuffmeier2023} or SPH methods \citep{bate2018}. Despite the different
   set-ups and physics used, all of the simulations stress that the inhomogeneous accretion via
   streamers is highly dependent on the environment and leads to instabilities, disk misalignment and
   even late replenishment of the disk \citep{kuffmeier2023}.

   While star formation can take place in molecular clouds in various different environments, the
   purpose of this paper is to explore the case of a collapsing core within a turbulent filament. The
   turbulent environment of the core is generated self-consistently by gas accretion onto the filament.
   As we do not resolve the full physics needed to explore the disk evolution, we concentrate on the
   analysis of the accretion onto the disk itself. The paper is organised as follows: first we present
   the numerical simulation and initial condition in \autoref{sec:setup}. Thereafter, we present the
   results in \autoref{sec:results} and its various subsections. We compare our simulation
   to other numerical studies in \autoref{sec:comp} and conclude in \autoref{sec:conc}.

%__________________________________________________________________

\section{Numerical set-up}
\label{sec:setup}

   The simulation was run with the code \textsc{ramses} by \citet{teyssier2002} using
   the MUSCL scheme \citep[Monotonic Upstream-Centred Scheme for Conservation Laws,][]{vanLeer1977},
   a second-order Godunov scheme for solving the discretised Euler equations in conservative form
   on a Cartesian grid. As Riemann solver we used the HLLC-Solver
   \citep[Harten-Lax-van Leer-Contact,][]{toro1994} together with the multidimensional MC slope limiter
   \citep[monotonized central-difference,][]{vanLeer1979}.

   We set up a forming filament in a radially converging flow as defined in \citet{heigl2020} as an
   large-scale initial condition. The simulation consists of an isothermal 3D box of size
   $0.1\,\mathrm{pc}$ with periodic boundary conditions in the filament direction, in this case the
   x-axis. The isothermal temperature is set to $10.0\,\mathrm{K}$ and a molecular weight of
   $\mu = 2.36$ which results in a sound speed of around $0.19\,\mathrm{km\,s^{-1}}$. We define a
   cylindrical inflow in a cylindrical shell at the edge of the box onto the central x-axis with a
   fixed density and radial velocity which are constantly reset at every time-step. In order to break
   the symmetry the density in the inflow region is varied on a cell-by-cell basis with a random value
   which varies around the mean with a maximum of 50\%. The inflow density and the density inside the
   domain is set to $3.92 \times 10^{-22}\,\mathrm{g\,cm^{-3}}$ which corresponds to a particle density
   of about $1.0 \times 10^2\,\mathrm{cm^{-3}}$ and the inflow velocity is set to Mach 6.0 for the
   isothermal gas which equals a velocity of $1.13\,\mathrm{km\,s^{-1}}$. The mass accretion rate per
   length onto the filament therefore corresponds to
   $\dot{M}/L = 8.4\,\mathrm{M_\sun\,pc^{-1}\,Myr^{-1}}$. This flow not only acts as an external
   pressure to confine the filament material but also drives turbulence inside the filament which
   quickly settles to a constant level. As shown in \citet{heigl2018}, the turbulence velocity is around
   $0.23\,\mathrm{km\,s^{-1}}$, or Mach 1.18, for an inflow velocity of Mach 6.0. This value is
   independent of the accretion flow density itself as well as the strength of its random
   variation as kinetic energy is converted to turbulent energy. However, the turbulent velocity does
   show an anti-correlation to the density profile with the centre of the filament showing lower values
   than the surface.

   The minimum resolution of the simulation is $256^3$ cells which is equivalent to
   $1.56 \times 10^{-3}\,\mathrm{pc}$ or about $323\,\mathrm{AU}$. However, we heavily make use of adaptive
   mesh refinement in order to resolve the scales down to the creation of an accretion disk with a
   maximum resolution of $16383^3$ cells corresponding to $2.44 \times 10^{-5}\,\mathrm{pc}$ or
   $5.04\,\mathrm{AU}$. In order to minimise dissipation across refinement levels and to resolve thin
   streamer structures in a large region around the forming disk, we employ an aggressive refinement
   criterion where the Truelove criterion for the maximum density has to be fulfilled by a factor of
   128 \citep{truelove1997}.

   As soon as the maximum refinement level reaches a density of
   $1.83 \times 10^{-13}\,\mathrm{g\,cm^{-3}}$ where the Jean's length is not resolved by at least four
   cells, a sink particle is placed in the centre if the region fulfills several conditions. In the
   standard version of \textsc{ramses} these conditions consist mainly of the gravitational potential
   dominating over the thermal and kinetic support. We also allow mass accretion onto the sinks by
   using the implemented Bondi accretion algorithm with the standard setting of the accretion radius
   of four cells. The amount and distribution of sink particles strongly depend on the formation
   conditions and the subsequent accretion. However, as this paper's main focus is the creation of
   streamers and as we do not include enough physics to properly describe the processes inside of the
   disk, we will explore the sink population in a future paper. In this study, the main purpose of the
   sink particles is to take the role of a mass sink in order to keep the simulation running. Doing so,
   their impact on the streamers are minimal.

%__________________________________________________________________

\section{Results}
\label{sec:results}

   \begin{figure}
     \centering
     \includegraphics[width=1.0\columnwidth]{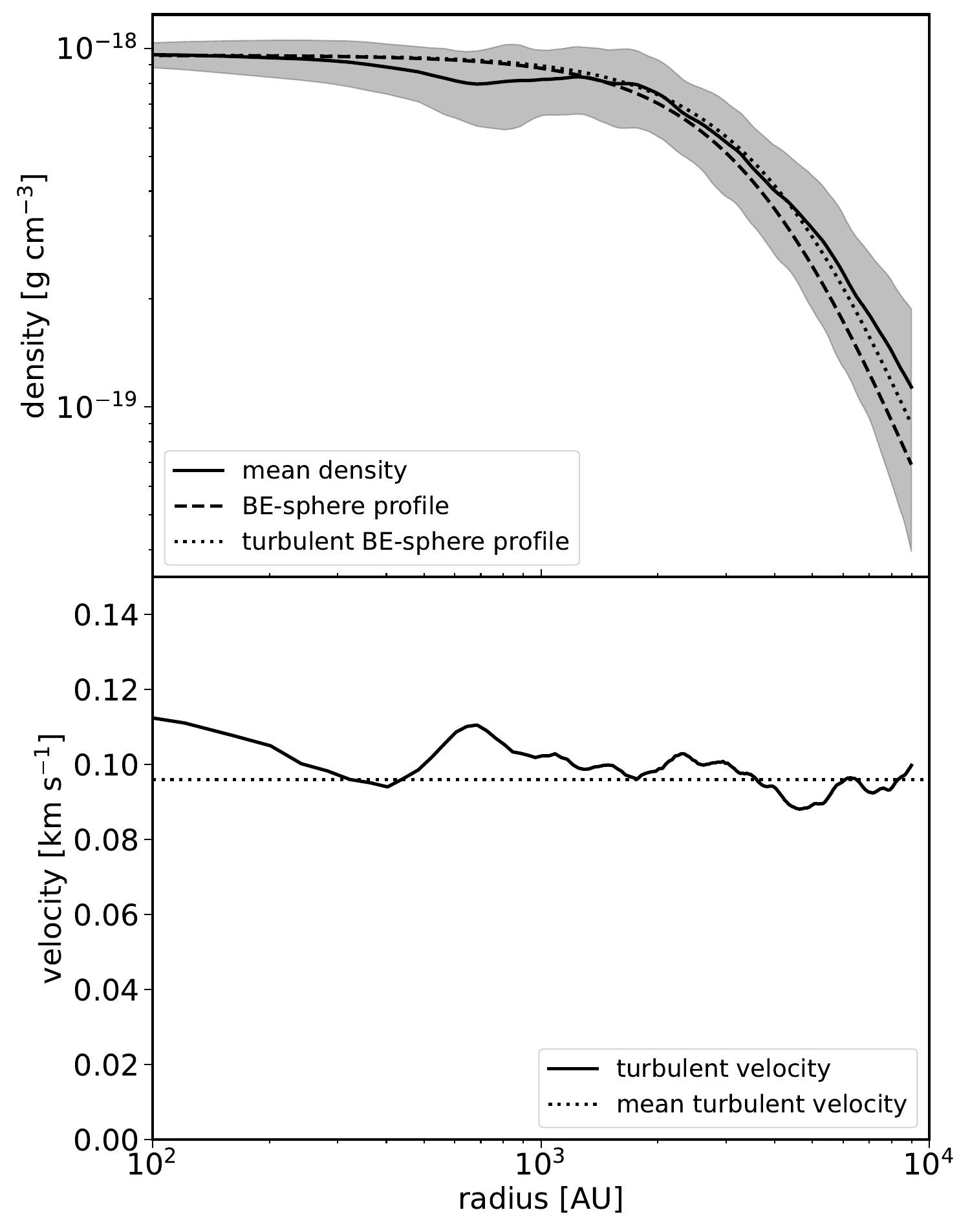}
     \caption{Radial density and velocity profiles of the forming
     core $93.2\,\mathrm{kyr}$ before its collapse. The upper panel shows
     the mean density profile as solid black line together with Bonner-Ebert
     sphere profiles where the dashed line is an
     isothermal model and the dotted line has an additional turbulent term.
     For this, we use the turbulent velocity and its mean given in the lower
     panel as solid and dotted black lines.}
     \label{fig:besphere}
   \end{figure}

   \begin{figure}
     \centering
     \includegraphics[width=1.0\columnwidth]{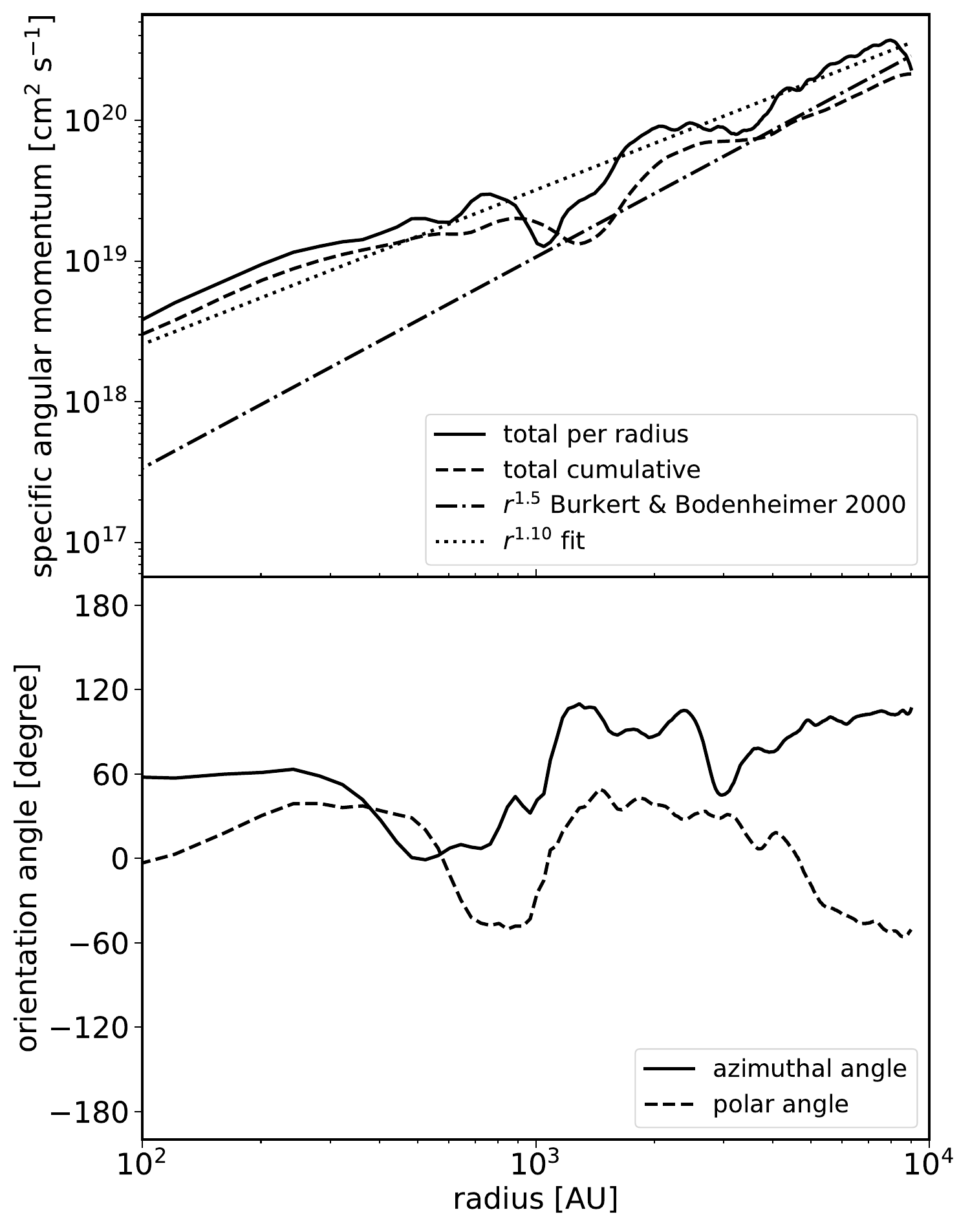}
     \caption{Total specific angular momentum of the forming core and its
     orientation $93.2\,\mathrm{kyr}$ before collapse.
     The upper panel shows the angular momentum of each radial shell as solid
     line as well as the cumulative total angular momentum up to that radius
     as dashed lines. We fit a slope to the former which shows that they both
     follow a close to linear trend given by the dotted line. We also plot the
     predicted slope for turbulent cores by \citet{burkert2000} as the
     dashed-dotted line. The lower panel shows the orientation of the total
     specific angular momentum vector of each radial shell.}
     \label{fig:ang}
   \end{figure}

   \subsection{Core collapse and disk formation}

   Over time a turbulent filament forms inside the box constrained by the accretion pressure of the
   radial flow. As soon as the filament's line-mass reaches close to critical values given by
   \begin{equation}
     \left(\frac{M}{L}\right)_\mathrm{crit} = \frac{2c_s^2}{G} \approx 1.06 \times 10^{16}\mathrm{\,g\,cm^{-1}} \approx 16.4\,\mathrm{M_{\sun}\,pc^{-1}},
   \end{equation}
   with $c_s$ being the isothermal sound speed and $G$ being the gravitational constant, a core begins
   to form as shown in \autoref{fig:core}. We do not impose any overdensity in the initial condition or
   the accretion flow and the location where the core forms is only determined by the initial seed for
   the random number generator. The core itself shows turbulent motions and substructure. Before its
   collapse, the core's density profile closely follows a Bonnor-Ebert distribution as shown in the upper
   panel of \autoref{fig:besphere}. The measured mean density in each radial bin is given by the solid
   line together with its standard deviation shown as the gray area. We compare it to the isothermal
   Bonnor-Ebert profile shown as the dashed line as well as the turbulent Bonnor-Ebert profile where we
   take into account the support of a turbulent pressure component to the sound speed as
   \begin{equation}
     \sigma_\mathrm{tot} = \sqrt{c_s^2+\sigma^2_\mathrm{1D}}
   \end{equation}
   where $\sigma_\mathrm{1D} = \sigma_\mathrm{3D}/\sqrt{3}$ is the turbulent velocity in one dimension
   which we calculate from the mean of the three-dimensional turbulent velocity of radial bins around
   the density maximum. This velocity is shown as solid line in the bottom panel of
   \autoref{fig:besphere}. One can see that the turbulent velocity is relatively constant throughout
   the Bonnor-Ebert sphere at a value with a mean of $0.096\,\mathrm{km\,s^{-1}}$ which we use for our
   turbulent Bonnor-Ebert profile in the upper panel given by the dotted line. The actual density seems
   to match the turbulent profile better, the error bars however are too large to distinguish between
   the isothermal and turbulent profile.

   \begin{figure}
     \centering
     \includegraphics[width=1.0\columnwidth]{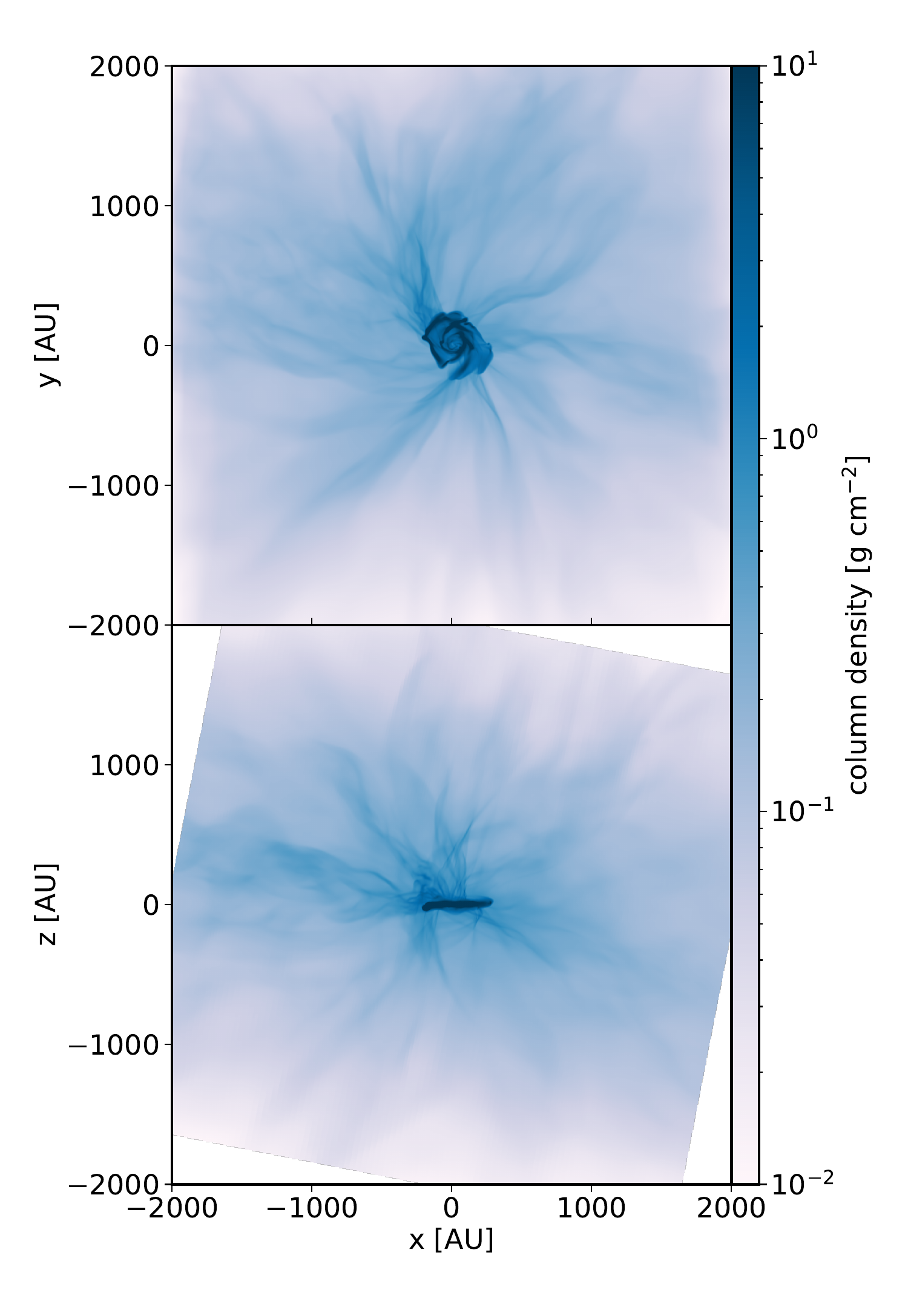}
     \caption{Face-on and edge-on density projections of the forming disk
     at around $24.32\,\mathrm{kyr}$ after sink formation. The box has a
     size of $4000\,\mathrm{AU}$ around the central sink and is rotated to
     align the disk. The colour bar is adjusted to highlight the denser
     streamers flowing onto the disk. The density enhancements are accreted
     from all directions, as well as from above and below the disk, and
     extend to several thousand astronomical units.}
     \label{fig:pro}
   \end{figure}

   We also check if there is an overall rotation of the core which provides the angular momentum for
   the formation of a disk and which can potentially influence the formation of streamers. For a core
   with solid body rotation there are analytic models which predict the accretion rate onto the disk
   as a function of time and radius \citep{ulrich1976, hueso2005}. Therefore, we also plot the total
   specific angular momentum in the upper panel of \autoref{fig:ang}. We calculate the total specific
   angular momentum as
   \begin{equation}
     \vec{j}_\mathrm{tot} = \frac{\sum \rho \vec{v} \times \vec{r}}{\sum \rho},
   \end{equation}
   where we use the density $\rho$, the velocity $\vec{v}$ and the radial position $\vec{r}$ of each
   cell. We determine the total specific angular momentum of each radial shell individually, shown as
   the solid line, as well as cumulatively for all cells within a given radius, shown by the dashed
   line. As the total angular momentum is non-zero, there is some bulk rotation which, however, is very
   small compared to overall turbulent motion.

   Both curves, the one calculated per radius and the cumulative one, seem to scale close to linearly
   with the radius as indicated by the fit to the former value given by the dotted line with a slope of
   $1.10\pm0.02$. This is due to the fact that the bulk velocity is relatively constant with radius,
   similar to the turbulent velocity. Observations of the cumulative total angular momentum rather
   suggest a scaling with radius to the power of 1.5 \citep{goodman1993, caselli2002,
   pirogov2003, chen2007, tobin2011, yen2015}. This scaling was proposed by \citet{burkert2000} as a
   consequence of sampling the turbulent velocity field at different scales and we also show it in the
   plot as dashed-dotted line. One can see, that it does not seem to match the total profiles as good
   as the linear fit. However, if we limit the fitted area to typical scales used in observations of
   above $1000\,\mathrm{AU}$, the slope is slightly steeper with a slope of $1.21\pm0.02$ which
   matches both scalings similarly well.

   While there seems to be some bulk rotation, the core shows no signs of solid body rotation which
   would suggest the total angular momentum to be proportional to the radius squared. Moreover, the
   bulk motion is not necessarily rotating in the same direction throughout the core as shown by the
   lower panel. Here, we plot the direction of the angular momentum vector for each radial shell.
   As one can see the azimuthal and polar angle are constantly changing throughout the core which
   proves that there is no ordered overall rotation.

   \begin{figure}
     \centering
     \includegraphics[width=1.0\columnwidth]{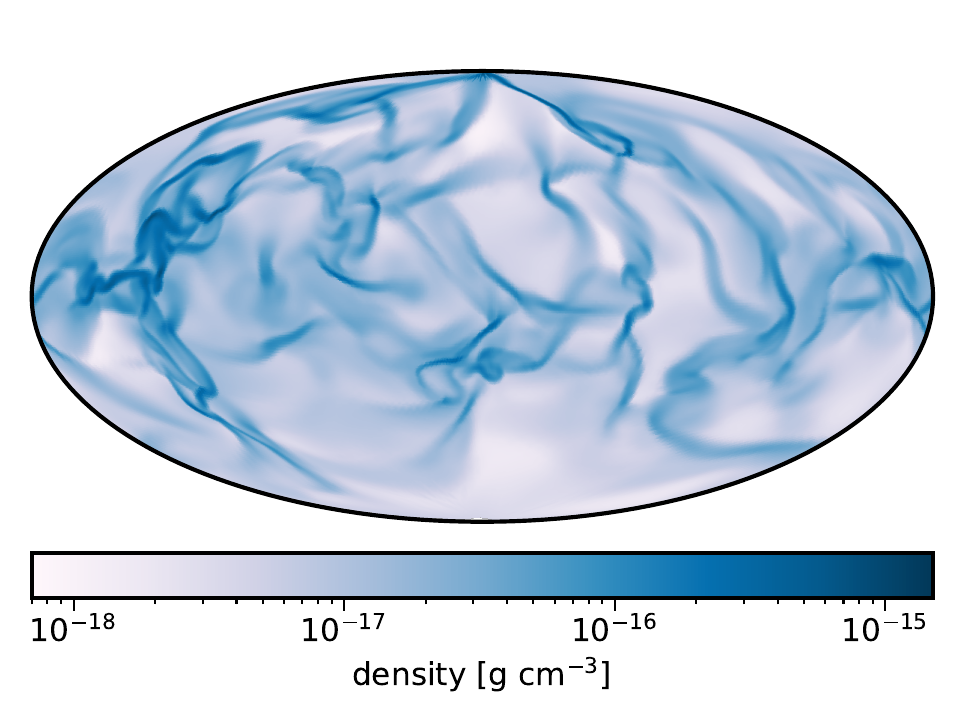}
     \caption{Hammer projection of a spherical shell of the density at
     a distance of $400\,\mathrm{AU}$ from the central sink. The
     overdensities are visible as elongated density enhancements.}
     \label{fig:rhopro}
   \end{figure}

   After a short time, the core begins to collapse and a sink particle forms in its centre. Over time,
   more and more material falls onto the sink particle and, as not all of it is accreted directly, a
   disk forms around it. This situation is shown in \autoref{fig:pro} which shows a face and edge-on
   projection of the density within a box of $4000\,\mathrm{AU}$ centred around the central sink
   particle. The sink has reached a mass of $0.68\,\mathrm{M_\sun}$ and the disk a mass of
   $0.09\,\mathrm{M_\sun}$ at $24.32\,\mathrm{kyr}$ after the sink formation. Depending on the column
   density contrast, one can clearly make out denser streams of material of lengths of up to several
   thousand AU flowing onto the disk. They do not necessarily settle to the disk plane before
   hitting the disk, but also seem to be accreted within the disk. For the further analysis of the
   streamer properties, we will concentrate on this snapshot.

   In order to investigate the geometry of the streamers, we cut a spherical shell out of the
   three-dimensional density distribution at a radius of $400\,\mathrm{AU}$ from the central sink
   particle and project it on a map using a Hammer projection as shown in \autoref{fig:rhopro}.
   One can see, that most of the area is filled by low density material of around
   $1.0 \times 10^{-18}\,\mathrm{g\,cm^{-3}}$ and that the denser material  of more than
   $1.0 \times 10^{-16}\,\mathrm{g\,cm^{-3}}$ forms concentrated, distinct structures. If the streamers
   would be filamentary by nature, they would appear as point-like density enhancements where they
   cross the shell. However, most of the higher density structures show a large extent in one direction
   which is typical of sheet-like structures. This also becomes apparent when looking at the density
   projection around the disk in different angles which show vastly different configurations of density
   enhancements which is caused by the projection of sheets along the line-of-sight. The sheet-like
   geometry of the overdensities was already noted in \citet{kuffmeier2017} and is consistent to large
   radii, however the density contrast also becomes weaker which can be seen below in
   \autoref{fig:veldivout} where we show the density maps for larger distances. We also note,
   that the Hammer projections are created from an interpolation on a grid of regular angle spacing.
   However, the statistical analysis shown in the next sections is performed on a Healpix map
   \citep{gorski1999} which uses an equal-area and iso-latitude pixelisation in order to circumvent an
   oversampling of high latitudes.

   \subsection{Infall velocities and accretion}

   \begin{figure}
     \centering
     \includegraphics[width=1.0\columnwidth]{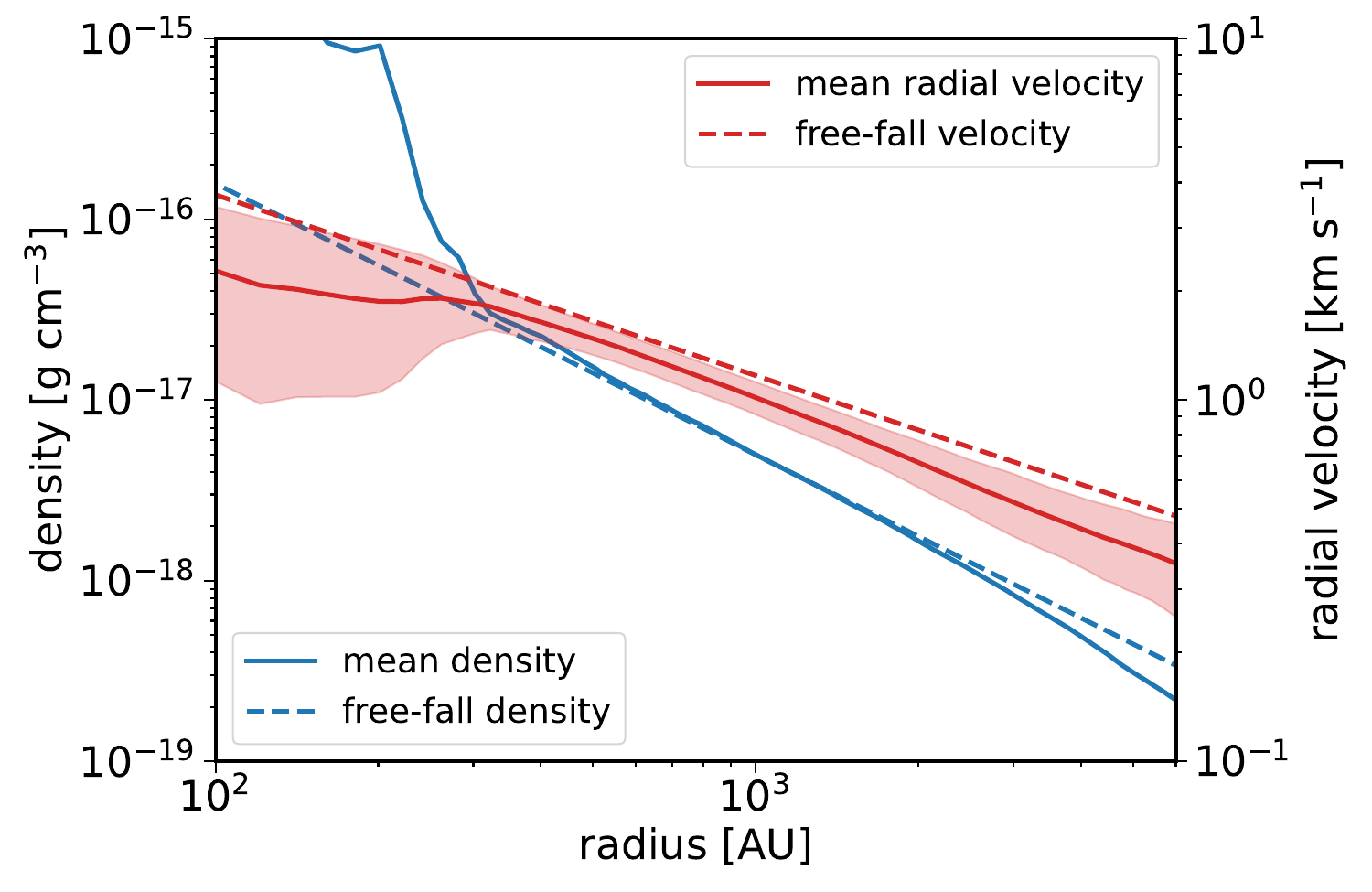}
     \caption{Radially averaged density and velocity profiles during collapse.
     The mean density is given by the solid blue and the mean radial velocity
     by the solid red line. We also show the expected density and velocity
     profile assuming a free-fall model given by the blue and red dashed lines,
     respectively.}
     \label{fig:radav}
   \end{figure}

   In order to measure the infall, we calculate the mean density and mean radial velocity at every
   radius which we show in \autoref{fig:radav} as the solid blue and red line together with the
   standard deviation of the radial velocity shown as the shaded red area. The standard deviation
   of the density is quite large, especially for small radii as the density varies over three orders
   of magnitude. Therefore, we do not show it in order to keep the plot legible. We compare the
   resulting profiles to the expected density and velocity of free-fall collapse given by the
   dashed blue and red line, respectively. The free-fall velocity is given by
   \begin{equation}
     v_\mathrm{ff} = \sqrt{\frac{2G\left(M_\ast+M_\mathrm{disk}\right)}{r}},
   \end{equation}
   with $M_\ast$ and $M_\mathrm{disk}$ being the mass of the central sink particle and the disk and $r$
   being the radius. The free-fall density is then calculated assuming that the mass accretion rate
   \begin{equation}
     \dot M = 4\pi r^2 \rho v_\mathrm{ff}
   \end{equation}
   is constant, at least for small radii, as there is no mass build-up in a radial shell. We also
   observe that the mass accretion rate is roughly constant at around
   $2.6 \times 10^{-5}\,\mathrm{M_\sun\,yr^{-1}}$ which we use to derive the density profile of
   \begin{equation}
     \rho_\mathrm{ff} = \frac{\dot M}{4\pi\sqrt{2GM_\ast r^3}}.
   \end{equation}
   One can see that both, the density and velocity, follow closely the free-fall model. For small radii
   below $280\,\mathrm{AU}$, the density shows a strong increase which is the accretion shock onto the
   outer radius of the disk where the infall velocity also reaches a maximum. For larger radii above
   $3000\,\mathrm{AU}$, we see that the free-fall model is not  a good description of the infall anymore
   which is where pressure effects start to play a larger role and where we have a transition to the
   filament material.

   \begin{figure}
     \centering
     \includegraphics[width=1.0\columnwidth]{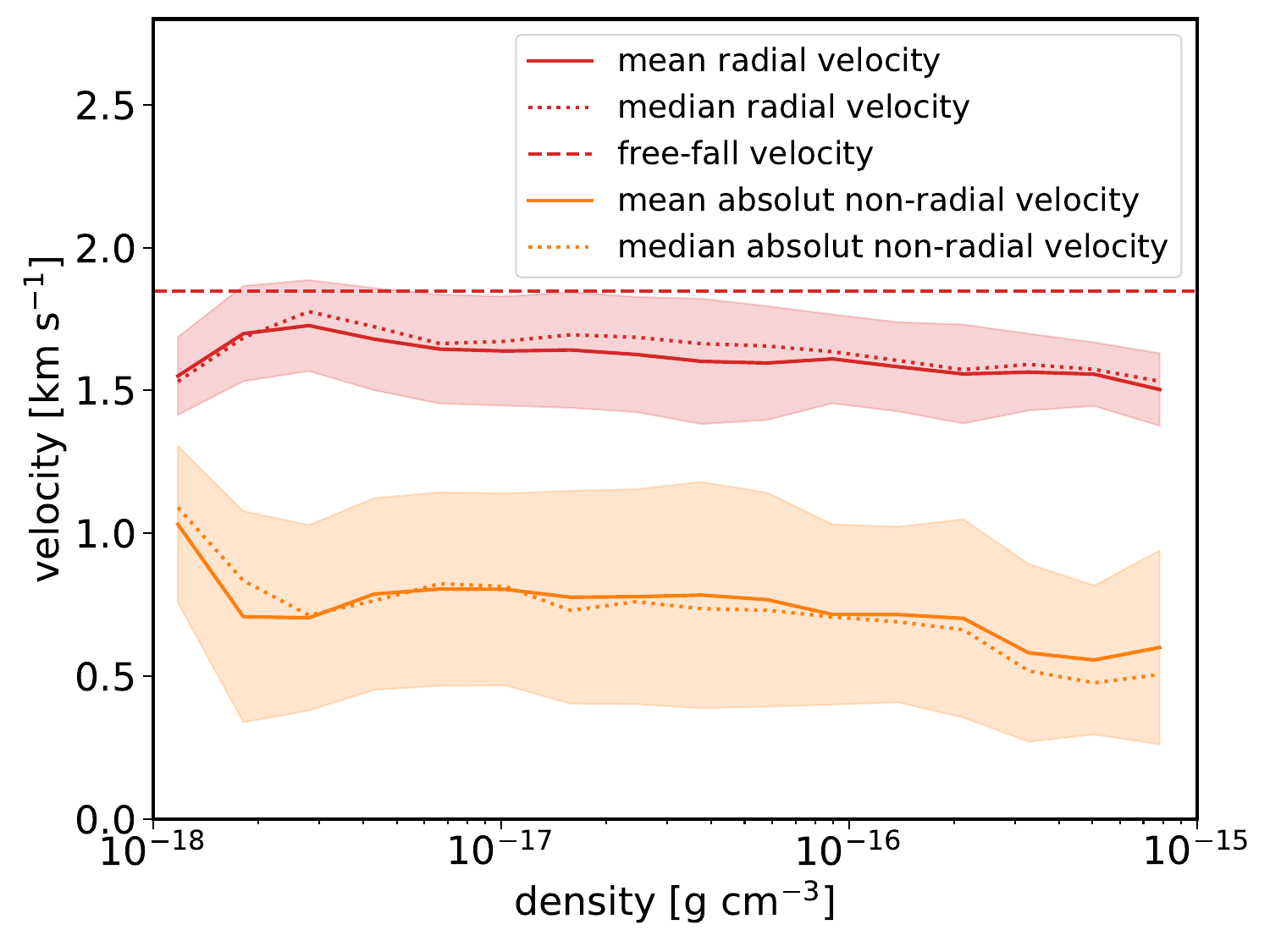}
     \caption{Velocity statistics as function of density at a distance of
     $400\,\mathrm{AU}$. The mean radial velocity is shown by the solid
     red and the mean absolute non-radial velocity by the solid orange
     line together with their standard deviation given by their respective
     shaded area. In addition, we also plot the median velocities given
     by the dotted lines and the free-fall velocity for the given
     central mass and distance shown by the dashed line.}
     \label{fig:vbins}
   \end{figure}

   An interesting aspect of the collapse is whether the density enhancements have a different radial and
   angular velocity compared to the low density regions. As material is concentrated into denser
   structures, its angular momentum could add up and force it onto more non-radial dominated orbits or
   the angular momentum could cancel out, forcing it onto more radial orbits. However, there does not
   seem to be a strong trend which we show in \autoref{fig:vbins}. At a distance of $400\,\mathrm{AU}$,
   we calculate the mean radial and absolute angular velocity within sixteen different logarithmic
   density bins, given by the solid red and orange line, respectively. In addition, we also calculate
   the median value as the distribution of the velocity within each density bin is not necessary
   Gaussian. As one can see the radial velocity is relatively independent of the density, despite it
   spanning three orders of magnitude. Overall, it is slightly lower but very close to the free-fall
   velocity given by the dashed red line of about $1.8\,\mathrm{km\,s^{-1}}$. The offset could be
   due to the reason that the disk at this distance cannot be approximated as a point-like gravitational
   source which should lead to a slightly lower free-fall velocity at small radii. There is a small
   decrease of the mean radial velocity in the very low density bin and a very shallow trend to lower
   velocities at the very high density bins. The reason for this at very low densities could be that
   the material can be more easily diverted onto non-radial orbits due to the gravitational attraction
   of the denser gas. We investigate this by looking at the mean absolute non-radial velocity
   component given in each density bin by the orange solid line in \autoref{fig:vbins} with a standard
   deviation depicted by the orange shaded area. The non-radial motions are much slower than the radial
   collapse with values of around $0.75\,\mathrm{km\,s^{-1}}$. And indeed, the lowest density bin shows
   a slight increase in non-radial motions indicating that there is a process forcing material onto
   non-radial orbits. We also see a general trend to slower velocities for larger densities which
   could be caused by the collimation of material into sheets where there is some canceling of angular
   momentum.

\begin{figure}
     \centering
     \includegraphics[width=1.0\columnwidth]{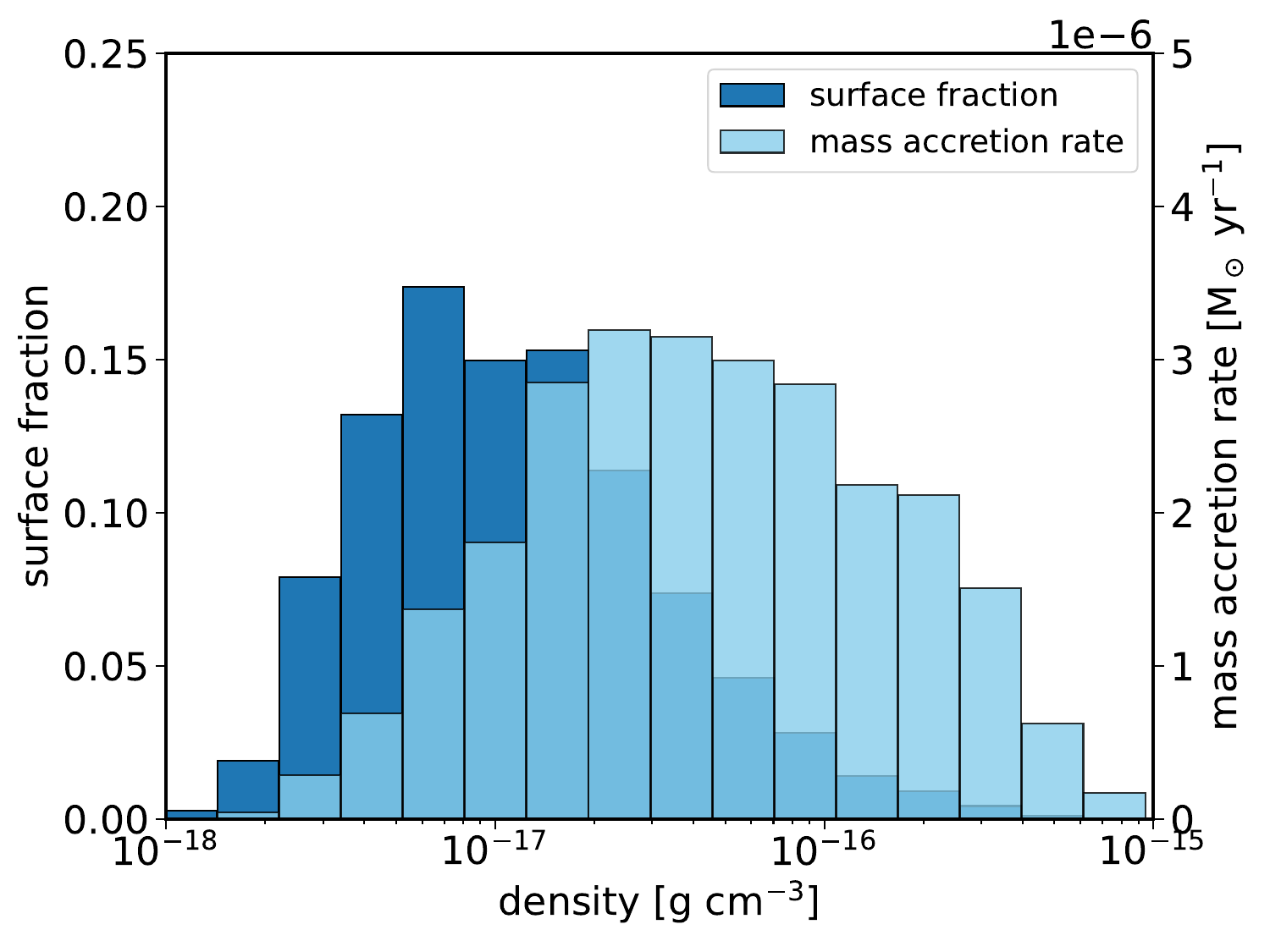}
     \caption{Mass accretion distribution as function of density at
     a distance of $400\,\mathrm{AU}$ from the central sink particle.
     The dark blue bars show the distribution of the surface fraction
     of each density bin and the light blue bars show their respective
     total accretion rate. As one can see, most mass is accreted at high
     densities and low surface fractions.}
     \label{fig:mdot}
   \end{figure}

   A consequence of a relatively uniform infall velocity is that the mass accretion rate scales with
   the density. This means that most of the mass which lands in the disk is accreted via the denser
   structures. We visualise this fact in \autoref{fig:mdot} where we plot the surface fraction in dark
   blue and the mass accretion rate in light blue in the same density bins as in \autoref{fig:vbins}.
   While most cells show a low density, with the maximum being at around
   $6.0 \times 10^{-18}\,\mathrm{g\,cm^{-3}}$, it is the high density cells above
   $2.0 \times 10^{-17}\,\mathrm{g\,cm^{-3}}$ where most mass is accreted. Focussing only on the very
   dense structure above $1.0 \times 10^{-16}\,\mathrm{g\,cm^{-3}}$ shows that close to half of the
   total mass accretion rate through the shell, namely $1.24 \times 10^{-5}\,\mathrm{M_\sun\,Myr^{-1}}$
   of $2.58 \times 10^{-5}\,\mathrm{M_\sun\,Myr^{-1}}$, is accreted in only a fraction of 0.1 of the
   total surface. Therefore, the dense accretion structures play an essential role in bringing in
   material to small scales. Moreover, given that \autoref{fig:vbins} shows that the angular velocity
   and thus the specific angular momentum is relatively constant in each density bin. Therefore, the
   accretion of absolute angular momentum shows the same distribution as the mass accretion rate.

\begin{figure}
     \centering
     \includegraphics[width=1.0\columnwidth]{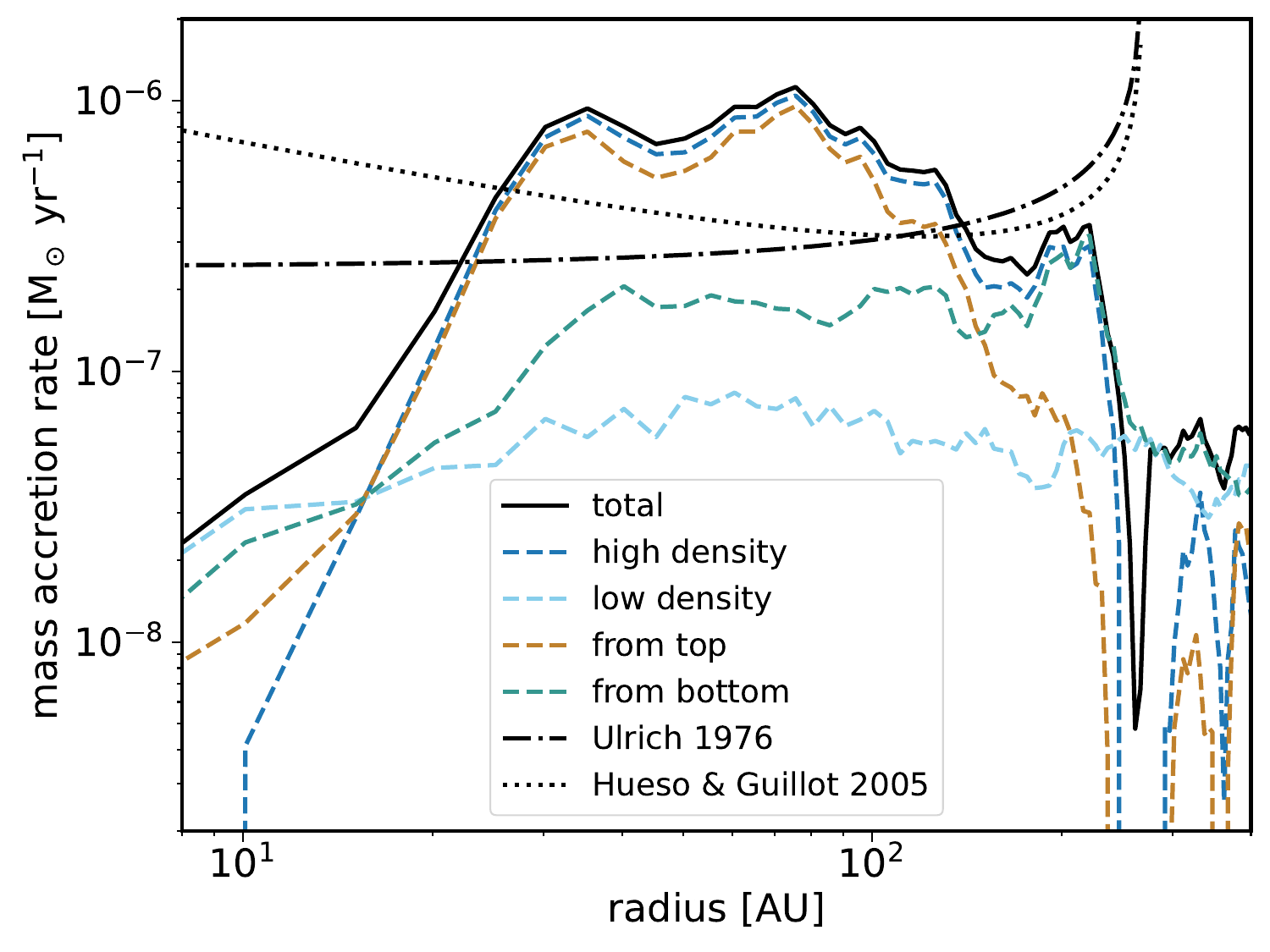}
     \caption{Total mass accretion onto the disk measured at 10 cells above
     and below the central plane given by the solid black line together
     with predictions from solid body rotation shown as dashed-dotted and
     dotted black lines. Compared to the models, we measure an increased mass
     accretion rate at intermediate radii. The total mass accretion rate is
     split into its side of origin as shown by the red and teal dashed lines
     and its density given by the light and dark blue dashed lines.}
     \label{fig:mdotdisk}
   \end{figure}

   Additionally, it is vital to understand where the material enters the disk as not all material is
   accreted at the outer radius. We measure the mass accretion rate onto the disk at each radius by
   azimuthally summing up the flux through the surfaces located 10 cells above and below the central
   plane using the velocity perpendicular to the disk. We show the resulting total mass accretion rate
   in \autoref{fig:mdotdisk} as the black solid line together with its side of origin, either the top or
   bottom face, as red and teal dashed lines. We also split the total mass accretion rate into low and
   high density material with a threshold density of $1.0 \times 10^{-16}\,\mathrm{g\,cm^{-3}}$ and plot
   the corresponding rates as light and dark blue dashed lines. As one can see, most mass is accreted
   between 30 and $100\,\mathrm{AU}$ with a steep drop towards smaller and larger radii up to the
   maximum extent of the disk of about $280\,\mathrm{AU}$. The integral over all radii equals the
   total mass accretion rate of $2.58 \times 10^{-5}\,\mathrm{M_\sun\,Myr^{-1}}$ which we already
   measured on larger scales. While this only represents one measurement for this particular point in
   time, we do observe this pattern of accretion onto intermediate radii to be relatively stable over
   timescales of several ten thousand years when normalising for the disk size.

   The accretion itself is highly asymmetrical which can be seen in the separate accretion rates for
   the different sides of the disk. While it is dominated at large radii by accretion coming from below
   the disk, the accretion at lower radii is mainly coming from above. This can also be observed in the
   projection plots of \autoref{fig:pro} and implies that the angular momentum is not distributed
   isotropically. Looking at the split-up between low and high density, one can also see that the bulk
   of the mass accretion consists of high density material. This could already be seen from
   \autoref{fig:mdot}, but shows that the density enhancements continue down onto the disk itself.
   Interestingly, the low density regime does not display the same strong tendency for a preferred
   accretion radius but shows a more distributed, albeit negligible, accretion with a similar rate at
   all radii.

   \begin{figure}
     \centering
     \includegraphics[width=1.0\columnwidth]{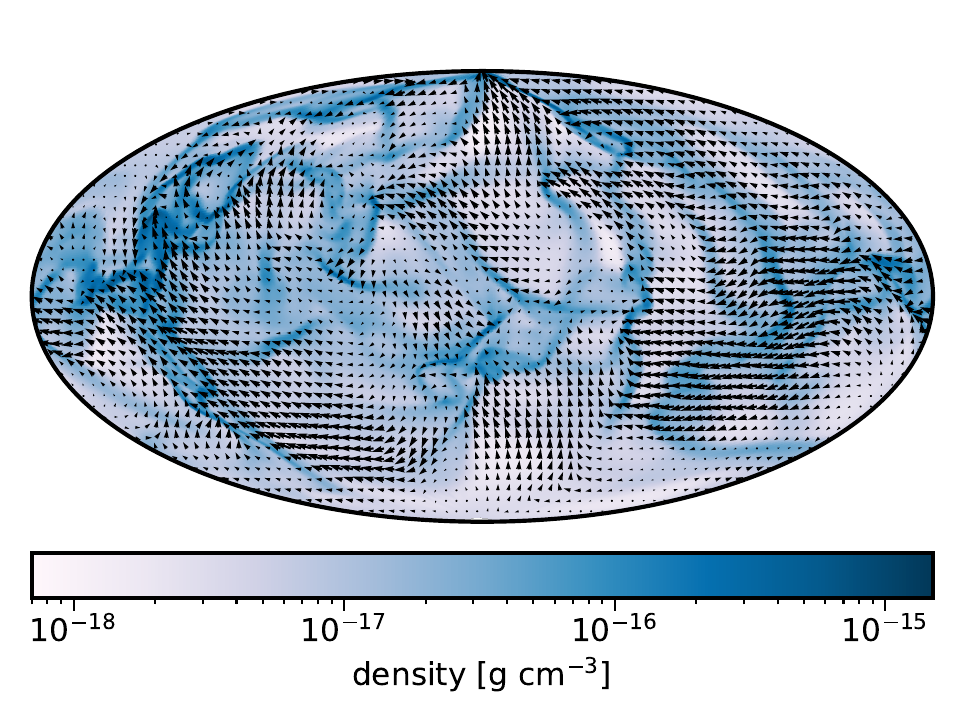}
     \caption{The same Hammer projection of the density given in
     \autoref{fig:rhopro} together with respective velocity vectors
     indicating the direction and magnitude of the motions
     perpendicular to the infall.}
     \label{fig:rhovel}
   \end{figure}

   We also include predictions from two different models for the mass accretion rate of a core under
   solid body rotation. Although we already showed that the initial core does not rotate as a solid
   body, this is the only model for which an analytic description of the mass accretion rate exist.
   The dashed-dotted line shows the model by \citet{ulrich1976} which assumes that every mass element
   of a radial shell hits the disk under a parabolic orbit according to its angular momentum with the
   central sink being in its focus. This leads to the so-called source function, the mass accretion
   rate per azimuthal surface:
   \begin{equation}
     S_U(r,t) = \frac{\dot M(t)}{4\pi r_c(t)^2}\left(\frac{r}{r_c(t)}\right)^{-1}\left(1-\frac{r}{r_c(t)}\right)^{-1/2}
   \end{equation}
   with $r_c$ being the centrifugal radius, the maximum impact radius, given by
   \begin{equation}
     r_c = \frac{\Omega^2 r_0^4}{GM_\ast}
   \end{equation}
   which depends on the shells angular speed $\Omega$ and its radius $r_0$. The dotted line uses the
   model by \citet{hueso2005} which assumes that the material is accreted directly onto its respective
   Keplerian radius leading to a slightly more centrally dominated source function:
   \begin{equation}
     S_H(r,t) = \frac{\dot M(t)}{8\pi r_c(t)^2}\left(\frac{r}{r_c(t)}\right)^{-3/2}\left(1-\sqrt{\frac{r}{r_c(t)}}\right)^{-1/2}.
   \end{equation}
   In principle, the mass accretion rate and the centrifugal radius both depend on time. However, in
   the case of free-fall, the mass accretion rate is constant and the evolution of centrifugal radius
   can be described analytically. In our case, where we do not start with solid body rotation, we
   cannot calculate the centrifugal radius as function of time. Therefore, we use the radial extent
   of the disk of $280\,\mathrm{AU}$ as an estimate for the models. As expected, the measured mass
   accretion rate deviates strongly from the solid body rotation case which predicts larger accretion
   rates towards the centre and close to the centrifugal radius. The deficit of material coming
   in at small radii is a consequence of the accreted material lacking very low angular momentum. For
   solid body rotation this is the material close to the rotational axis. However, due to the turbulent
   initial condition, there is no ordered rotation and therefore not much gas with very low angular
   momentum. The same argument is valid inversely for the mass accretion rate at very large radii. For
   solid body rotation this is given by the material in the plane of rotation. However, in our case,
   the maximum angular momentum is limited by the random motion of the gas which is statistically
   more unlikely to reach large values consistently.

   As the core is collapsing essentially in free-fall and each shell is accreted consecutively with
   larger and larger angular momentum, the mean main accretion radius should be set by the total
   angular momentum of the shell where the gas originated. As not all mass of the shell is
   combined into one accretion stream, there should be some spread in angular momentum of the
   individual overdensities. However, this spread should be around the total angular momentum of the
   original shell. We can test this by determining the original radius in two different ways. On the
   one hand, we can compare the angular momentum of the current mean main accretion to the initial angular
   momentum distribution. One the other hand, we can calculate the distance the current main accretion
   travelled under the assumption of free-fall collapse.

   To estimate the total angular momentum we use the mass accretion peak at $80\,\mathrm{AU}$ and
   the angular velocity of the dense gas of around $0.6\,\mathrm{km s^{-1}}$ to get a rough value of
   $7.0 \times 10^{19}\,\mathrm{cm^2\,s^{-1}}$.
   Comparing this value to \autoref{fig:ang} this means that the material is coming from a shell of
   roughly two to three thousand AU. The distance the gas would travel due to the free-fall time in
   about $25\,\mathrm{kyr}$ also equals around two thousand AU. This means that here the main
   accretion radius indeed seems to be set by the shell's angular momentum. If the angular momentum
   of accretion streams is always distributed around the shells original total angular momentum,
   the mean main accretion radius should expand over time as its value is increasing with further
   distance. However, because the angular momentum of each individual shell is not always aligned to
   the other shells, accretion is also able to heavily disrupt the disk. We also observe this in our
   simulation where the disk size can change quite drastically over time.

   \begin{figure}
     \centering
     \includegraphics[width=1.0\columnwidth]{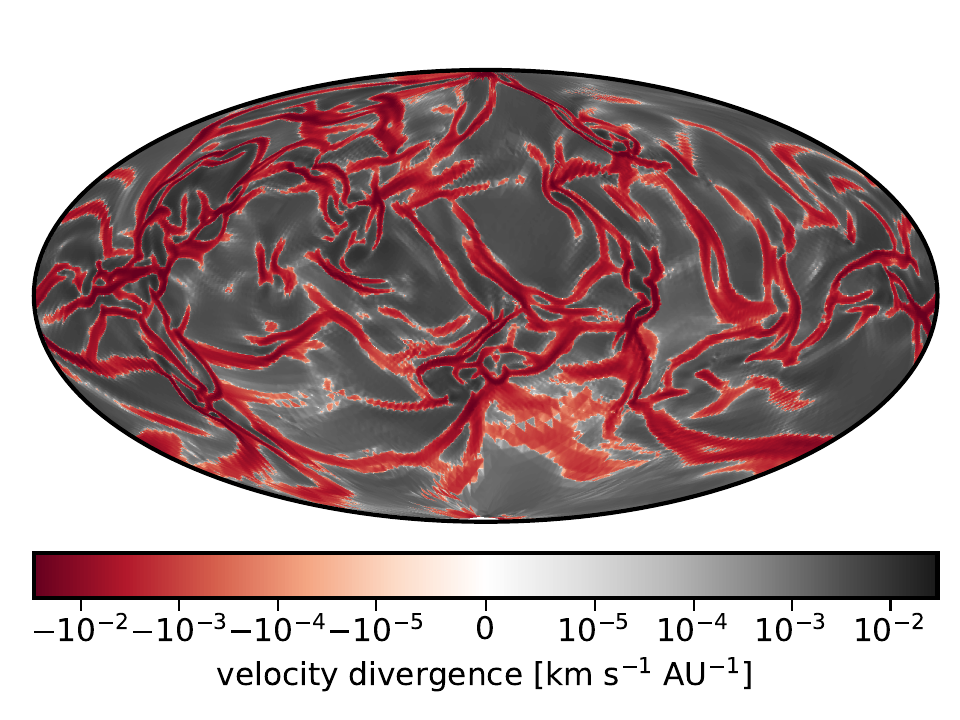}
     \caption{Hammer projection of the velocity divergence of the
     non-radial velocity components at $400\,\mathrm{AU}$ distance from the
     central sink particle. Areas in red show structures where
     material is compressed due to the velocity field.}
     \label{fig:veldiv}
   \end{figure}

   \subsection{Residual velocities}

   \begin{figure*}
     \centering
     \begin{subfigure}{\columnwidth}
         \includegraphics[width=\columnwidth]{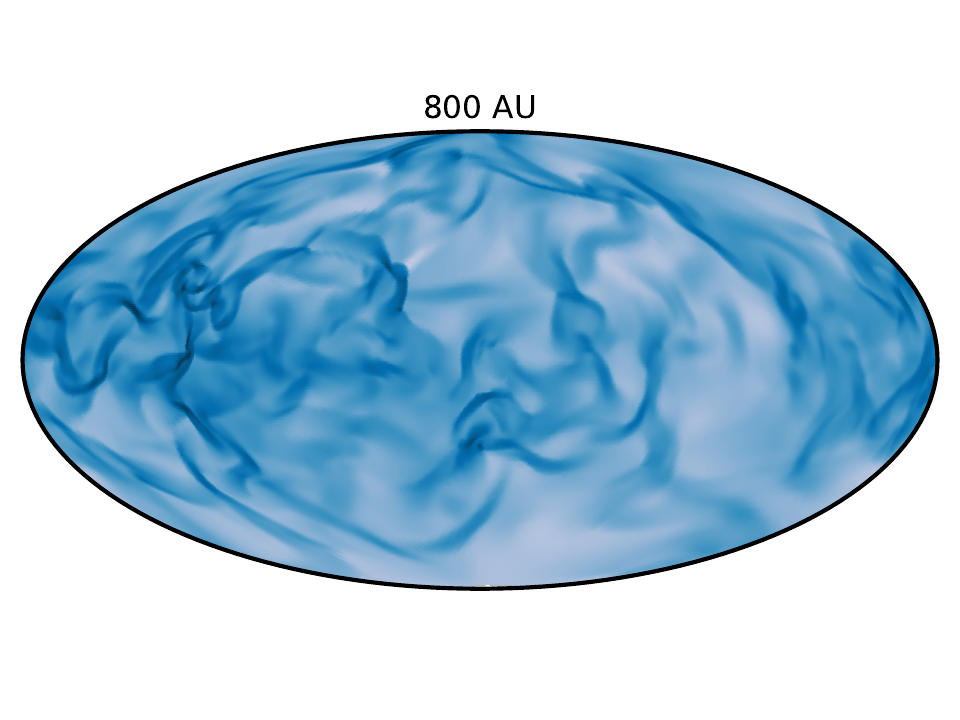}
     \end{subfigure}
     \begin{subfigure}{\columnwidth}
         \includegraphics[width=\columnwidth]{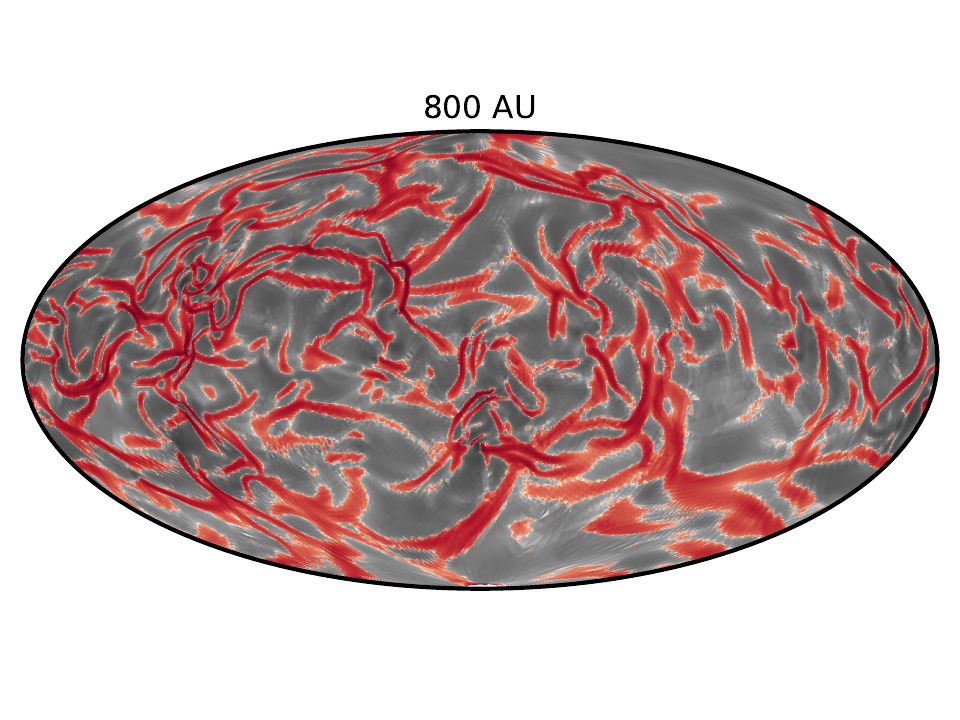}
     \end{subfigure}
     \begin{subfigure}{\columnwidth}
         \includegraphics[width=\columnwidth]{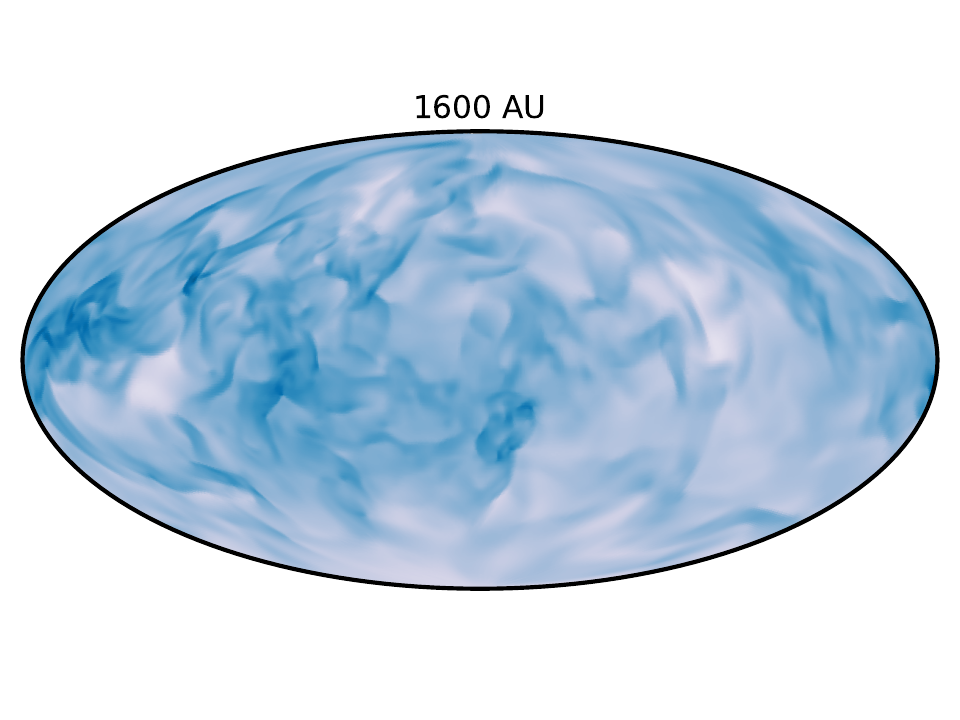}
     \end{subfigure}
     \begin{subfigure}{\columnwidth}
         \includegraphics[width=\textwidth]{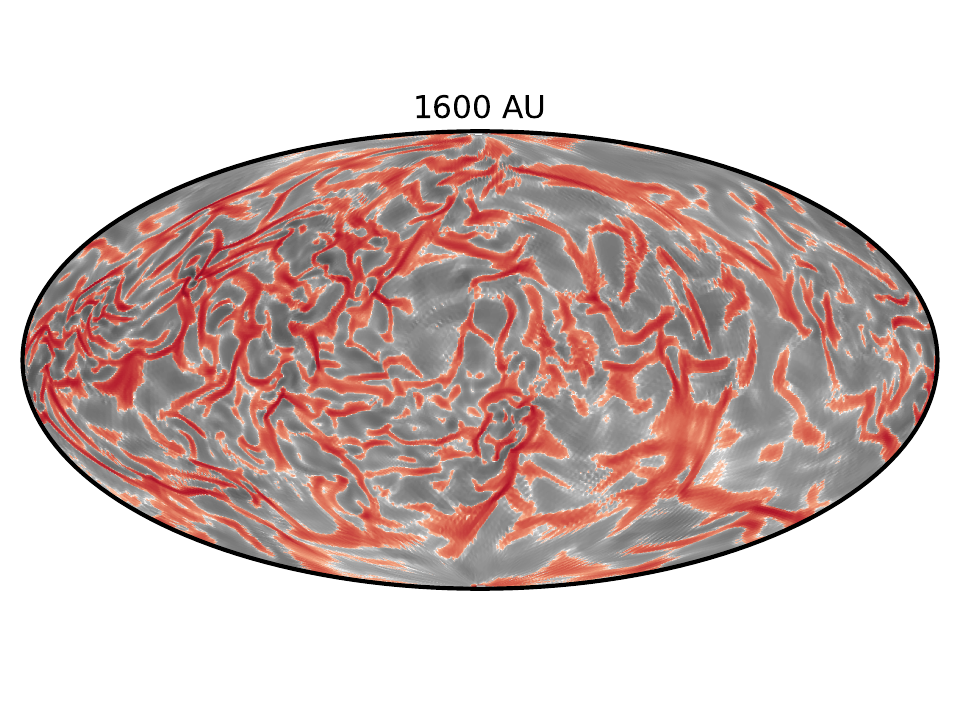}
     \end{subfigure}
     \begin{subfigure}{\columnwidth}
         \includegraphics[width=\columnwidth]{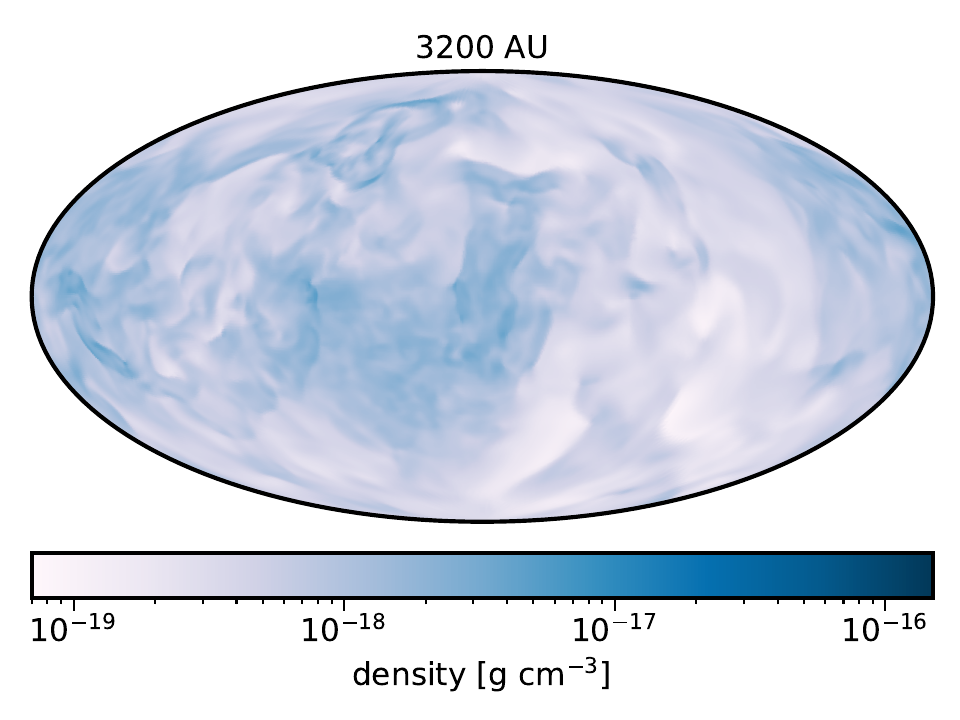}
     \end{subfigure}
     \begin{subfigure}{\columnwidth}
         \includegraphics[width=\textwidth]{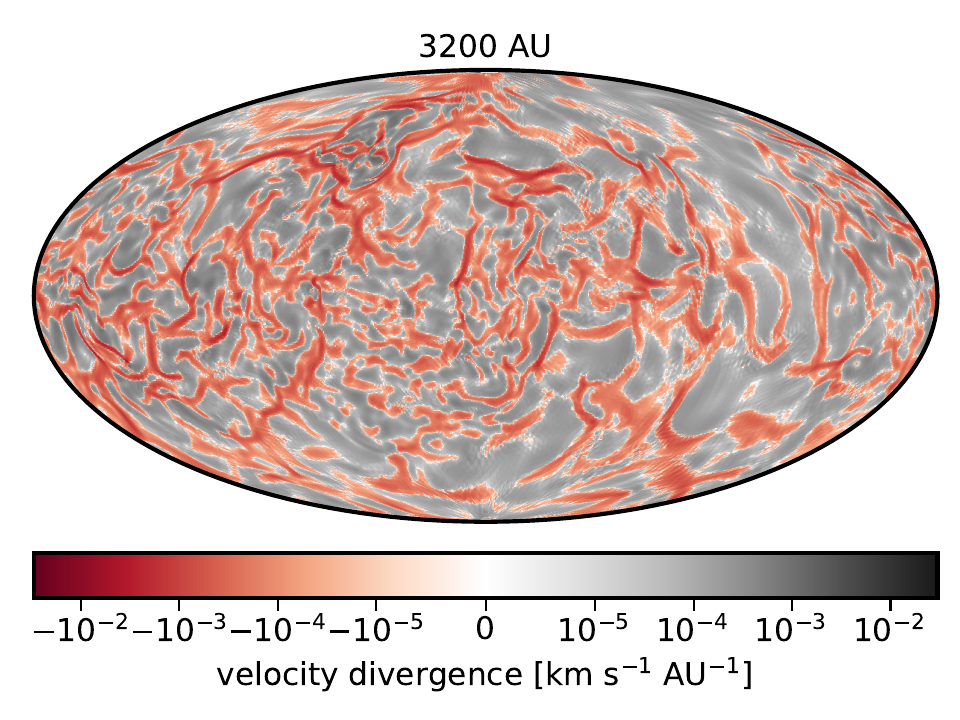}
     \end{subfigure}
     \caption{Hammer projections of the density and velocity divergence at
     different distances for the same time step. Form top to bottom we
     show further and further radii. As one can see, during infall the gas
     forms less but more condensed structures the closer it gets to the
     central sink particle.}
     \label{fig:veldivout}
   \end{figure*}

   \begin{figure}
     \centering
     \includegraphics[width=1.0\columnwidth]{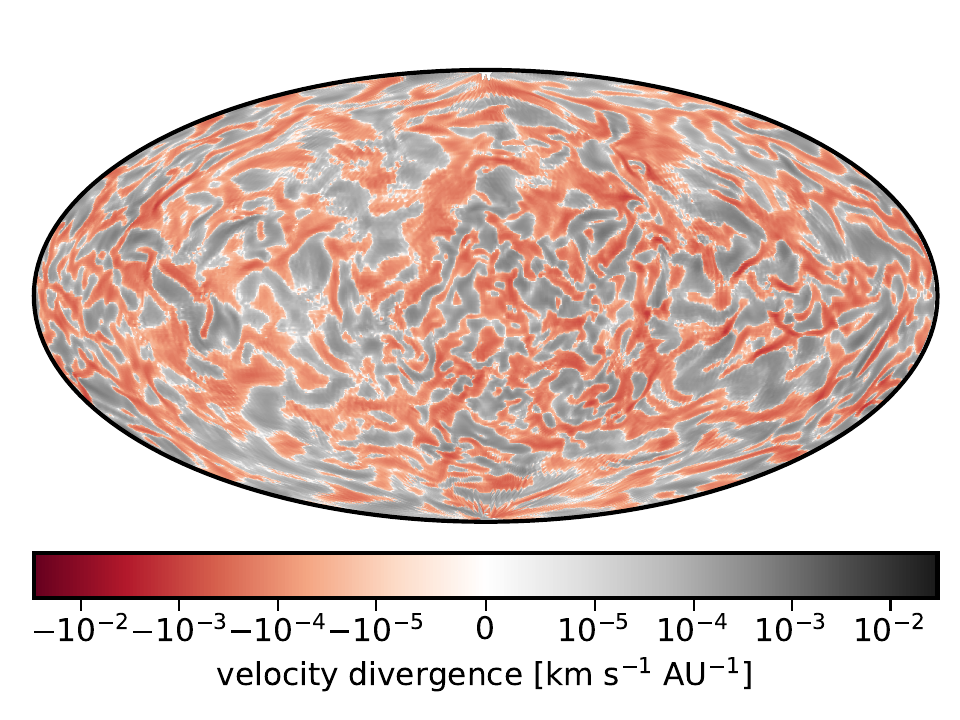}
     \caption{Velocity divergence at $3200\,\mathrm{AU}$ at around
     $93.2\,\mathrm{kyr}$ before sink formation. The velocity field is
     dominated by turbulent motions and already shows the typical
     structures visible during collapse in \autoref{fig:veldivout}
     which lead to the formation of the dense accretion sheets.}
     \label{fig:veldivpre}
   \end{figure}

   From the analysis of the radial velocity it is clear that essentially all material is falling onto
   the central sink in a free-fall manner. This only leaves the residual motions perpendicular to the
   infall as cause of the collimation of material into denser sheets. Therefore, we analyse the residual
   velocities in more detail. In \autoref{fig:rhovel}, we show the same map as in \autoref{fig:rhopro}
   with the perpendicular motions displayed on top where we use the azimuthal and polar velocity
   components to determine the direction and magnitude of the flow. As one can see, the density
   enhancements coincide either with strong changes in magnitude of the flow or where two flows
   converge. This means that the main locations of the formation of accretion sheets are given by where
   gas with different angular momentum collides and forms shock-like density enhancements.

   In order to emphasise this point, we also calculate the velocity divergence of the residual motions
   which we show in \autoref{fig:veldiv}. Note that this map only takes into consideration the azimuthal
   and polar velocities and neglects the radial velocity component. Negative values of the velocity
   divergence shown in red are areas where material is being compressed and positive values in gray
   show areas where material is being diluted. Comparing the map to the density map of
   \autoref{fig:rhovel}, one can clearly see that the areas of compression coincide with overdensities
   which shows again that it is the flows of the residual motions during free-fall which bring material
   together into denser accretion sheets.

   During infall, there will be a wide range of angular momentum. Gas with more eccentric orbits will
   at some point cross the orbits of the more radially dominated orbits of the denser accretion sheets.
   This means the density enhancements will accumulate more and more material along their way, leading
   to fewer but even denser structures. We show this effect in \autoref{fig:veldivout}. From top to
   bottom we plot the density and velocity divergence at further and further distances. As one can see,
   as the size and contrast of structures visible in the density goes down the farther out we go, the
   residual motions define more and more small scale convergence zones. Even at large distances, where
   radial collapse has not yet collimated a lot of material and the residual motions are still
   dominated by the turbulent velocity field of the filament, one can already observe convergence zones.
   This suggests that the initial seeds of overdensities are given by the turbulent motions on large
   scale. In order to verify that it is the turbulent velocity field which creates this pattern, we also
   plot the velocity divergence around the core of \autoref{fig:core} at $93.2\,\mathrm{kyr}$ before the
   collapse in \autoref{fig:veldivpre}. Here, the forming core only consists of an overdensity of a
   factor of two. As one can see, even at this early stage where their is no radial collapse yet, a
   very similar pattern is visible. This means that the initial turbulence on large scales indeed sets
   the formation pattern of the dense accretion sheets.

   \begin{figure}
     \centering
     \includegraphics[width=1.0\columnwidth]{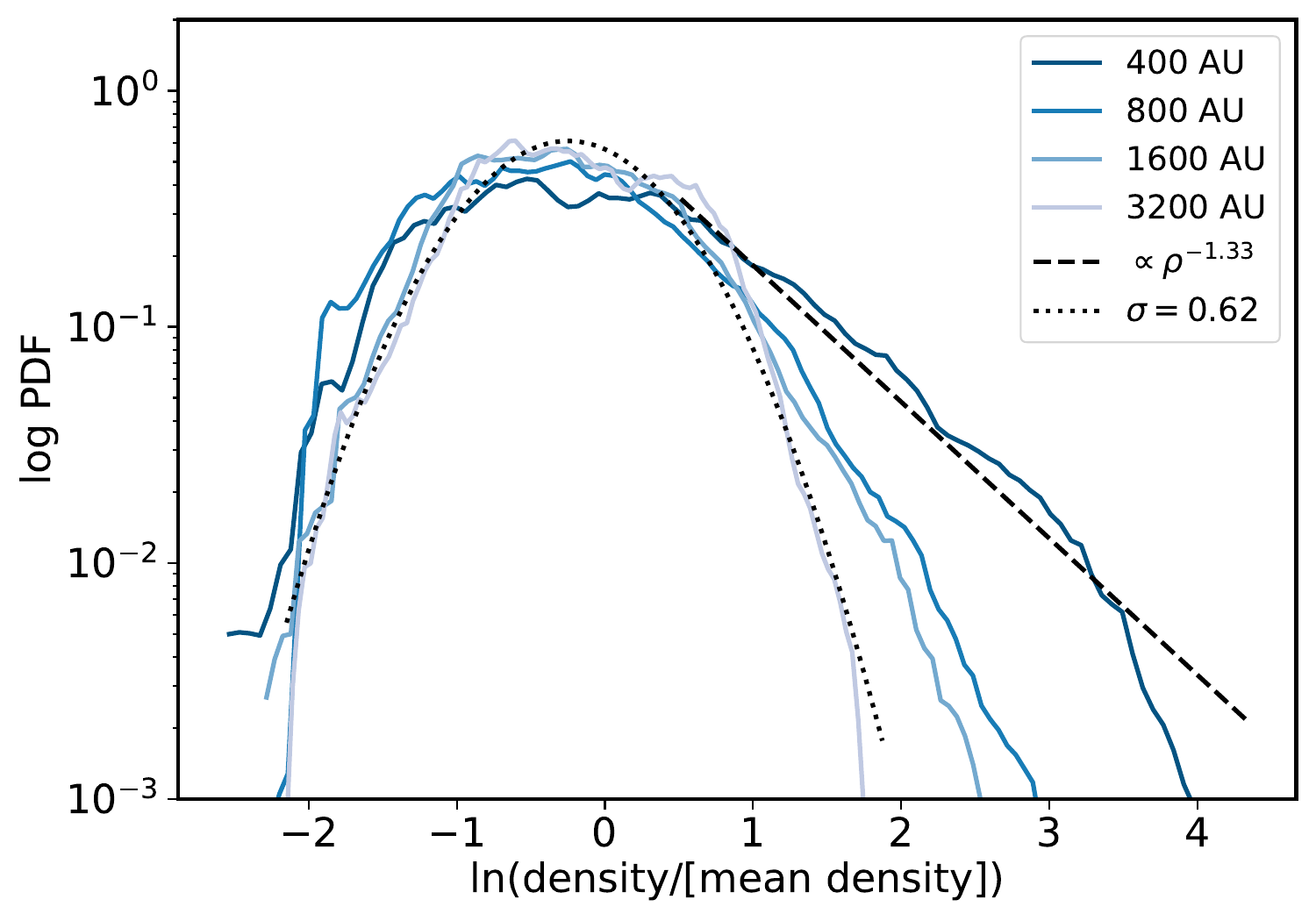}
     \caption{Density probability distribution of shells at different
     distances as given by the legend. The distribution follows a
     log-normal function with gas closer to the central sink particle
     showing the development of a power-law.}
     \label{fig:dpdf}
   \end{figure}

   One can also measure the collimation of material during infall in the density probability distribution
   at different distances as shown in \autoref{fig:dpdf}. From light to dark blue, we plot the density
   probability distribution of spherical shells starting from large distances down to close to the disk.
   One can see a systematic trend where the density at large distances follows a log-normal distribution
   with a dispersion of $\sigma=0.62\pm0.02$, whereas if one goes closer to the disk, one can see the
   development of a power-law tail. In that regard, the evolution from larger to smaller radii follows
   a similar process as the density in a turbulent box under gravitational collapse. The longer gravity is
   able to compress  the material, the more pronounced the power-law tail becomes. This suggests the same
   process is happening during the infall onto the central sink. Indeed, if we fit a power-law to the
   high-density tail we get a similar but slightly smaller index of $-1.33\pm0.03$ as in simulations of
   turbulent boxes which vary around -1.5 up to -2.5 \citep{kritsuk2011, federrath2013}. This suggests
   that gravity plays an essential role for the evolution of the overdensities which we will explore in
   the following subsection.

   \begin{figure}
     \centering
     \includegraphics[width=1.0\columnwidth]{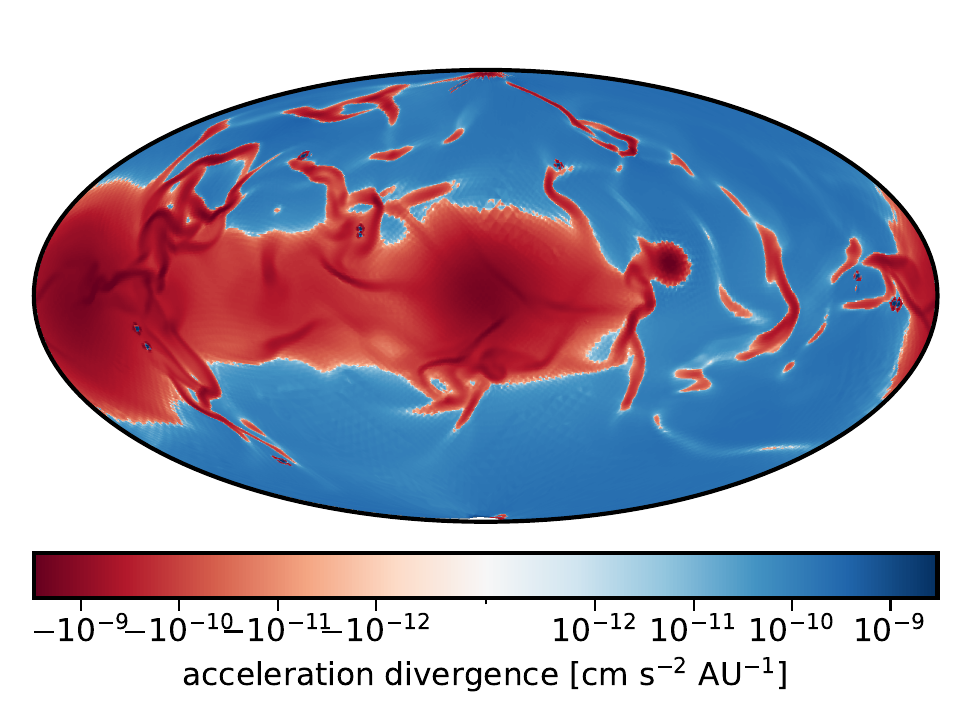}
     \caption{Hammer projection of the acceleration divergence at
     $400\,\mathrm{AU}$ from the central sink particle. Negative values
     show areas where material is gravitationally compressed and positive
     values where it is pulled apart. The gravitational compression of the
     disk is visible as the large distributed area together with the most
     massive accretion structures.}
     \label{fig:accdiv400}
   \end{figure}

   \begin{figure}
     \centering
     \includegraphics[width=1.0\columnwidth]{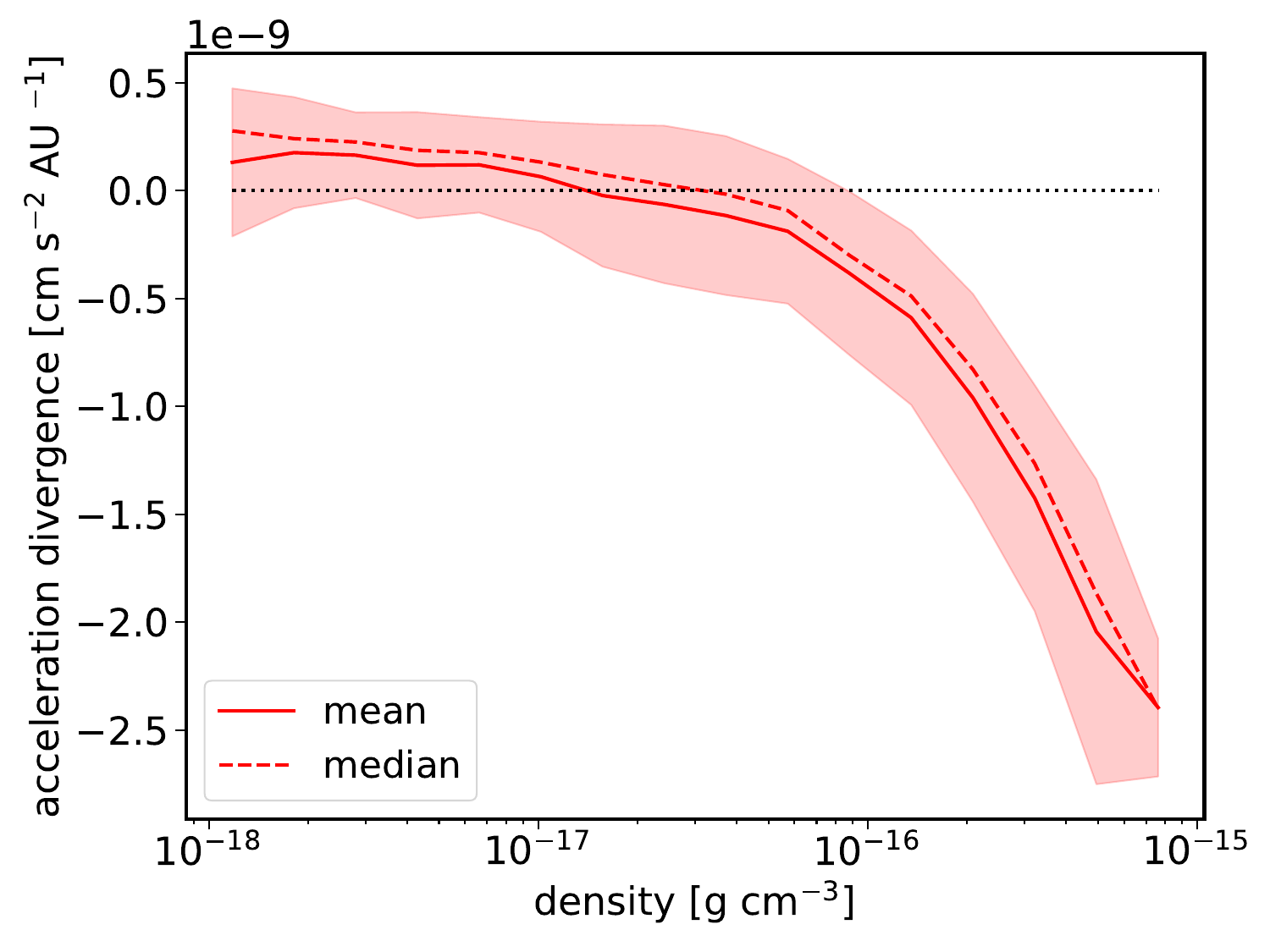}
     \caption{Mean acceleration divergence per density bin at $400\,\mathrm{AU}$
     distance of the central sink together with its standard deviation.
     While for low densities the divergence is positive and small, one
     can see that the gravitational compression dominates for densities
     larger than $1.0 \times 10^{-16}\,\mathrm{g\,cm^{-3}}$ where there
     is a steep drop to large negative values.}
     \label{fig:accbins}
   \end{figure}

   \subsection{Are the overdensities gravitationally dominated?}

   \begin{figure}
     \centering
     \includegraphics[width=1.0\columnwidth]{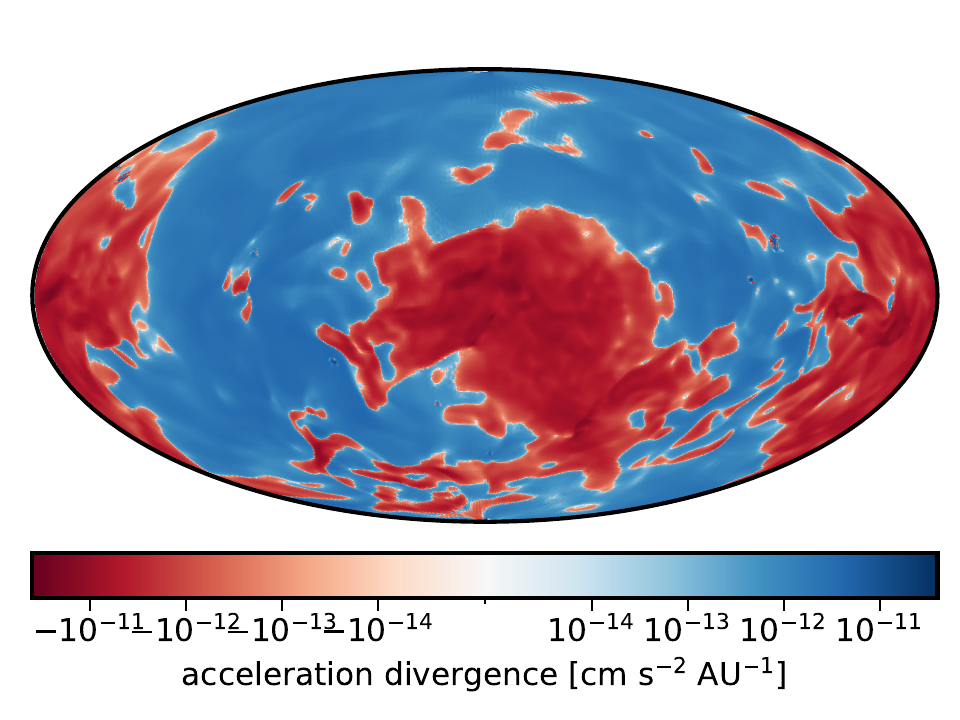}
     \caption{Hammer projection of the acceleration divergence at
     $3200\,\mathrm{AU}$ from the central sink particle. The centre
     and the border of the image point in the direction of the filament.
     The plot shows the large-scale gravitational acceleration onto the
     central axis due to the filament potential.}
     \label{fig:accdiv3200}
   \end{figure}

   In order to explore the question if the overdensities gravitationally dominate their local
   environment, we calculate the divergence of the acceleration field analogously to the velocity
   field and show the results in \autoref{fig:accdiv400}. Negative values coloured in red show areas
   where material is gravitationally compressed while positive values coloured in blue show areas where
   material is being gravitationally pulled apart. Very prominent is the gravitational attraction
   of the disk itself, pulling material into the central plane. It dominates over the gravitational
   pull of the streamers except for the area towards the right hand side of the viewing direction.
   Here it is interrupted by the gravitational field of one of the densest streamers and a close-by
   sink particle which pull the gas stronger into their direction, turning their vicinity blue.
   Compared to \autoref{fig:rhopro}, one can see that there is a one-to-one correspondence of the
   highest overdensities to areas with large negative divergence of the acceleration. However, not
   all structures are able to gravitationally pull material together. Some overdensities visible in
   the density map have no correspondence in the acceleration divergence map. We investigate this
   further in \autoref{fig:accbins} where we show the mean and median acceleration divergence
   as function of the density using the same bins as in \autoref{fig:vbins}. As one can see, only
   the very dense structures above around $1.0 \times 10^{-16}\,\mathrm{g\,cm^{-3}}$ show a
   significant negative acceleration divergence which means only the densest fraction of the
   overdensities are able to gravitationally concentrate material at this radius.

   This pattern remains the same going to larger distances until the gravitational pull onto the
   central axis of the filament dominates as can be seen in \autoref{fig:accdiv3200}. Compared, to
   the small scale velocity divergence at large distances visible in \autoref{fig:veldivout}, the
   missing gravitational divergence of small scale structure means that overdensities
   are mainly initialised by the velocity field. During infall, they are collimated and accrete
   enough material to gravitationally dominate their vicinity and gather even more material. This
   explains the increase in non-radial orbits of low density material which can be easily pulled
   towards the dense streamers seen in \autoref{fig:vbins} as well as the development of a power-law
   tail in the density probability function observed in \autoref{fig:dpdf}.

%______________________________________________________________

\section{Comparison to related work}
\label{sec:comp}

   Our main finding that the accretion is set by the large-scale environment agrees with other
   simulations such as \citet{kuffmeier2017}, \citet{bate2018} and \citet{lebreuilly2021}. We show
   that the formation of overdensities occurs naturally in a turbulent environment due to the initial
   velocity field. We find that this has a strong influence on where mass is accreted onto the disk.
   However, this effect may depend strongly on the included physics. For example, the simulation
   of \citet{lee2021} which includes magnetic fields finds a centrally concentrated mass accretion.

   Although the general outcome of overdensity formation during collapse is robust for different
   initial conditions, we made several simplifications in our simulations. We conducted this work in
   order to set a baseline to compare future parameter studies to. Of major importance would be the
   inclusion of magnetic fields. Depending on the strength of the magnetic field, it can have a
   significant impact on the gas dynamics, potentially leading to a reduction of overdensities.
   Magnetic fields also have a strong effect particularly on the sizes of disks due to magnetic braking
   and outflows \citep{allen2003, galli2006, price2007, hennebelle2008, duffin2009, commercon2010,
   seifried2011, seifried2012out} which typically leads to rather small disk radii. While we
   deliberately did not analyse the disk properties in this study, we do find a rather large disk.
   However, it is still in a very dynamic stage where the size can vary quite drastically. Moreover,
   depending on dust properties and cosmic-ray ionisation rates, non-ideal MHD effects can reduce the
   impact of strong magnetic fields, which again leads to more frequent disk formation and larger disk
   sizes \citep{shu2006, tsukamoto2015ohm, masson2016, zhao2016, zhao2018, wurster&bate2019}.
   Furthermore, there is also evidence that the effect of magnetic breaking can be reduced by
   non-idealised initial conditions where turbulence plays an essential role, not necessarily for
   magnetic reconnection, but by misaligning the angular momentum from the magnetic field vector
   \citep{hennebelle2009, seifried2012, seifried2013, joos2013, li2013, wurster2019, hirano2020}. A
   comparison of the accretion including MHD effects will be the focus of a future study.

   In addition, while many studies include a barotropic equation-of-state in order to mimic thermal
   effects of the disk above the opacity limit, radiative feedback from the star has been shown to
   affect the temperature structure and therefore the fragmentation of the disk \citep{commercon2010,
   tsukamoto2015rad}. Moreover, thermal pressure is able of driving outflows even in the complete
   absence of magnetic fields \citep{bate2011, bate2014}. To which extent radiative feedback has an
   effect on overdensities in the accretion flow has to be explored in future simulations

%______________________________________________________________

\section{Conclusions}
\label{sec:conc}

   In this study, we simulated the collapse of a core which forms self-consistently within a turbulent
   filament in order to analyse the origin and properties of overdensities in the accretion flow. We
   analysed the morphological, kinematic and gravitational properties of the core before
   and during the collapse in great detail and are able to draw the following conclusions:

   \begin{enumerate}
      \item Overdensities in the accretion flow are created naturally as a consequence of the initial
         turbulent velocity field. This velocity field is set on large scales and is already present
         before the collapse.
      \item Instead of having a filamentary morphology, we find that the overdensities in our
         simulation are rather sheet-like where the observable accretion pattern depends on the
         line-of-sight projection. This morphology follows from the shock-like convergence of residual
         motions during collapse.
      \item The collapse is essentially in free-fall with no significant difference in the radial
         velocity of the high and low-density gas despite it spanning several orders of magnitude.
         We only see a slight tendency for lower angular velocities, e.g. more direct radial infall,
         of the high density gas and higher angular velocity, e.g. more eccentric infall, of the low
         density gas.
      \item The mass accretion onto the disk differs drastically from an idealised initial condition of
         a core with solid body rotation. This is due to the fact that the core inherits a turbulent
         velocity distribution from its environment with no generally ordered rotation.
      \item The turbulent velocity field defines a log-normal probability density function at large
         radii which develops a power-law tail at smaller radii, similar to the gravitational collapse
         in a turbulent box.
      \item Both, the increase in angular velocity at low densities and the development of a power-law
         tail at large densities, are a consequence of the gravitational pull of the most massive
         overdensities which are able to gravitationally dominate their local environment. This leads
         to more and more collimated structures over time during the infall.
   \end{enumerate}

\begin{acknowledgements}

   This research was supported by the Excellence Cluster ORIGINS which is funded by the Deutsche
   Forschungsgemeinschaft (DFG, German Research Foundation) under Germany’s Excellence Strategy
   - EXC-2094 - 390783311.
\end{acknowledgements}

\bibliographystyle{aa}
\bibliography{Diskpaper}

\begin{thebibliography}{93}
\expandafter\ifx\csname natexlab\endcsname\relax\def\natexlab#1{#1}\fi

\bibitem[{{Allen} {et~al.}(2003){Allen}, {Li}, \& {Shu}}]{allen2003}
{Allen}, A., {Li}, Z.-Y., \& {Shu}, F.~H. 2003, \apj, 599, 363

\bibitem[{{Andr{\'e}} {et~al.}(2014){Andr{\'e}}, {Di Francesco},
  {Ward-Thompson}, {Inutsuka}, {Pudritz}, \& {Pineda}}]{andre2014}
{Andr{\'e}}, P., {Di Francesco}, J., {Ward-Thompson}, D., {et~al.} 2014, in
  Protostars and Planets VI, ed. H.~{Beuther}, R.~S. {Klessen}, C.~P.
  {Dullemond}, \& T.~{Henning}, 27--51

\bibitem[{{Ansdell} {et~al.}(2017){Ansdell}, {Williams}, {Manara}, {Miotello},
  {Facchini}, {van der Marel}, {Testi}, \& {van Dishoeck}}]{ansdell2017}
{Ansdell}, M., {Williams}, J.~P., {Manara}, C.~F., {et~al.} 2017, \aj, 153, 240

\bibitem[{{Bae} {et~al.}(2015){Bae}, {Hartmann}, \& {Zhu}}]{bae2015}
{Bae}, J., {Hartmann}, L., \& {Zhu}, Z. 2015, \apj, 805, 15

\bibitem[{{Bate}(2011)}]{bate2011}
{Bate}, M.~R. 2011, \mnras, 417, 2036

\bibitem[{{Bate}(2018)}]{bate2018}
{Bate}, M.~R. 2018, \mnras, 475, 5618

\bibitem[{{Bate} {et~al.}(2014){Bate}, {Tricco}, \& {Price}}]{bate2014}
{Bate}, M.~R., {Tricco}, T.~S., \& {Price}, D.~J. 2014, \mnras, 437, 77

\bibitem[{{Burkert} \& {Bodenheimer}(2000)}]{burkert2000}
{Burkert}, A. \& {Bodenheimer}, P. 2000, \apj, 543, 822

\bibitem[{{Cabedo} {et~al.}(2021){Cabedo}, {Maury}, {Girart}, \&
  {Padovani}}]{cabedo2021}
{Cabedo}, V., {Maury}, A., {Girart}, J.~M., \& {Padovani}, M. 2021, \aap, 653,
  A166

\bibitem[{{Caselli} {et~al.}(2002){Caselli}, {Benson}, {Myers}, \&
  {Tafalla}}]{caselli2002}
{Caselli}, P., {Benson}, P.~J., {Myers}, P.~C., \& {Tafalla}, M. 2002, \apj,
  572, 238

\bibitem[{{Chen} {et~al.}(2007){Chen}, {Launhardt}, \& {Henning}}]{chen2007}
{Chen}, X., {Launhardt}, R., \& {Henning}, T. 2007, \apj, 669, 1058

\bibitem[{{Commer{\c{c}}on} {et~al.}(2010){Commer{\c{c}}on}, {Hennebelle},
  {Audit}, {Chabrier}, \& {Teyssier}}]{commercon2010}
{Commer{\c{c}}on}, B., {Hennebelle}, P., {Audit}, E., {Chabrier}, G., \&
  {Teyssier}, R. 2010, \aap, 510, L3

\bibitem[{{Dr{\k{a}}{\.z}kowska} \& {Dullemond}(2018)}]{drazkowska2018}
{Dr{\k{a}}{\.z}kowska}, J. \& {Dullemond}, C.~P. 2018, \aap, 614, A62

\bibitem[{{Duffin} \& {Pudritz}(2009)}]{duffin2009}
{Duffin}, D.~F. \& {Pudritz}, R.~E. 2009, \apjl, 706, L46

\bibitem[{{Federrath} \& {Klessen}(2013)}]{federrath2013}
{Federrath}, C. \& {Klessen}, R.~S. 2013, \apj, 763, 51

\bibitem[{{Foster} \& {Chevalier}(1993)}]{foster1993}
{Foster}, P.~N. \& {Chevalier}, R.~A. 1993, \apj, 416, 303

\bibitem[{{Galli} {et~al.}(2006){Galli}, {Lizano}, {Shu}, \&
  {Allen}}]{galli2006}
{Galli}, D., {Lizano}, S., {Shu}, F.~H., \& {Allen}, A. 2006, \apj, 647, 374

\bibitem[{{Goodman} {et~al.}(1993){Goodman}, {Benson}, {Fuller}, \&
  {Myers}}]{goodman1993}
{Goodman}, A.~A., {Benson}, P.~J., {Fuller}, G.~A., \& {Myers}, P.~C. 1993,
  \apj, 406, 528

\bibitem[{{Gorski} {et~al.}(1999){Gorski}, {Wandelt}, {Hansen}, {Hivon}, \&
  {Banday}}]{gorski1999}
{Gorski}, K.~M., {Wandelt}, B.~D., {Hansen}, F.~K., {Hivon}, E., \& {Banday},
  A.~J. 1999, arXiv e-prints, astro

\bibitem[{{Greaves} \& {Rice}(2011)}]{greaves2011}
{Greaves}, J.~S. \& {Rice}, W.~K.~M. 2011, \mnras, 412, L88

\bibitem[{{Harsono} {et~al.}(2018){Harsono}, {Bjerkeli}, {van der Wiel},
  {Ramsey}, {Maud}, {Kristensen}, \& {J{\o}rgensen}}]{harsono2018}
{Harsono}, D., {Bjerkeli}, P., {van der Wiel}, M. H.~D., {et~al.} 2018, Nature
  Astronomy, 2, 646

\bibitem[{{Heigl} {et~al.}(2018){Heigl}, {Burkert}, \&
  {Gritschneder}}]{heigl2018}
{Heigl}, S., {Burkert}, A., \& {Gritschneder}, M. 2018, \mnras, 474, 4881

\bibitem[{{Heigl} {et~al.}(2020){Heigl}, {Gritschneder}, \&
  {Burkert}}]{heigl2020}
{Heigl}, S., {Gritschneder}, M., \& {Burkert}, A. 2020, \mnras, 495, 758

\bibitem[{{Hennebelle} \& {Ciardi}(2009)}]{hennebelle2009}
{Hennebelle}, P. \& {Ciardi}, A. 2009, \aap, 506, L29

\bibitem[{{Hennebelle} {et~al.}(2020){Hennebelle}, {Commer{\c{c}}on}, {Lee}, \&
  {Charnoz}}]{hennebelle2020}
{Hennebelle}, P., {Commer{\c{c}}on}, B., {Lee}, Y.-N., \& {Charnoz}, S. 2020,
  \aap, 635, A67

\bibitem[{{Hennebelle} \& {Fromang}(2008)}]{hennebelle2008}
{Hennebelle}, P. \& {Fromang}, S. 2008, \aap, 477, 9

\bibitem[{{Hirano} {et~al.}(2020){Hirano}, {Tsukamoto}, {Basu}, \&
  {Machida}}]{hirano2020}
{Hirano}, S., {Tsukamoto}, Y., {Basu}, S., \& {Machida}, M.~N. 2020, \apj, 898,
  118

\bibitem[{{Hsieh} {et~al.}(2019){Hsieh}, {Hirano}, {Belloche}, {Lee}, {Aso}, \&
  {Lai}}]{hsieh2019}
{Hsieh}, T.-H., {Hirano}, N., {Belloche}, A., {et~al.} 2019, \apj, 871, 100

\bibitem[{{Hsieh} {et~al.}(2023){Hsieh}, {Segura-Cox}, {Pineda}, {Caselli},
  {Bouscasse}, {Neri}, {Lopez-Sepulcre}, {Valdivia-Mena}, {Maureira},
  {Henning}, {Smirnov-Pinchukov}, {Semenov}, {M{\"o}ller}, {Cunningham},
  {Fuente}, {Marino}, {Dutrey}, {Tafalla}, {Chapillon}, {Ceccarelli}, \&
  {Zhao}}]{hsieh2023}
{Hsieh}, T.~H., {Segura-Cox}, D.~M., {Pineda}, J.~E., {et~al.} 2023, \aap, 669,
  A137

\bibitem[{{Hueso} \& {Guillot}(2005)}]{hueso2005}
{Hueso}, R. \& {Guillot}, T. 2005, \aap, 442, 703

\bibitem[{{Joos} {et~al.}(2013){Joos}, {Hennebelle}, {Ciardi}, \&
  {Fromang}}]{joos2013}
{Joos}, M., {Hennebelle}, P., {Ciardi}, A., \& {Fromang}, S. 2013, \aap, 554,
  A17

\bibitem[{{Kido} {et~al.}(2023){Kido}, {Takakuwa}, {Saigo}, {Ohashi}, {Tobin},
  {J{\o}rgensen}, {Aikawa}, {Aso}, {Encalada}, {Flores}, {Gavino}, {de
  Gregorio-Monsalvo}, {Han}, {Hirano}, {Koch}, {Kwon}, {Lai}, {Lee}, {Lee},
  {Li}, {Lin}, {Looney}, {Mori}, {Narayanan}, {Plunkett}, {Phuong}, {(Insa
  Choi)}, {Santamar{\'\i}a-Miranda}, {Sharma}, {Sheehan}, {Thieme}, {Tomida},
  {van't Hoff}, {Williams}, {Yamato}, \& {Yen}}]{kido2023}
{Kido}, M., {Takakuwa}, S., {Saigo}, K., {et~al.} 2023, \apj, 953, 190

\bibitem[{{Kritsuk} {et~al.}(2011){Kritsuk}, {Norman}, \&
  {Wagner}}]{kritsuk2011}
{Kritsuk}, A.~G., {Norman}, M.~L., \& {Wagner}, R. 2011, \apjl, 727, L20

\bibitem[{{Kruijer} {et~al.}(2017){Kruijer}, {Burkhardt}, {Budde}, \&
  {Kleine}}]{kruijer2017}
{Kruijer}, T.~S., {Burkhardt}, C., {Budde}, G., \& {Kleine}, T. 2017,
  Proceedings of the National Academy of Science, 114, 6712

\bibitem[{{Kruijer} {et~al.}(2014){Kruijer}, {Touboul}, {Fischer-G{\"o}dde},
  {Bermingham}, {Walker}, \& {Kleine}}]{kruijer2014}
{Kruijer}, T.~S., {Touboul}, M., {Fischer-G{\"o}dde}, M., {et~al.} 2014,
  Science, 344, 1150

\bibitem[{{Kuffmeier} {et~al.}(2019){Kuffmeier}, {Calcutt}, \&
  {Kristensen}}]{kuffmeier2019}
{Kuffmeier}, M., {Calcutt}, H., \& {Kristensen}, L.~E. 2019, \aap, 628, A112

\bibitem[{{Kuffmeier} {et~al.}(2021){Kuffmeier}, {Dullemond}, {Reissl}, \&
  {Goicovic}}]{kuffmeier2021}
{Kuffmeier}, M., {Dullemond}, C.~P., {Reissl}, S., \& {Goicovic}, F.~G. 2021,
  \aap, 656, A161

\bibitem[{{Kuffmeier} {et~al.}(2018){Kuffmeier}, {Frimann}, {Jensen}, \&
  {Haugb{\o}lle}}]{kuffmeier2018}
{Kuffmeier}, M., {Frimann}, S., {Jensen}, S.~S., \& {Haugb{\o}lle}, T. 2018,
  \mnras, 475, 2642

\bibitem[{{Kuffmeier} {et~al.}(2017){Kuffmeier}, {Haugb{\o}lle}, \&
  {Nordlund}}]{kuffmeier2017}
{Kuffmeier}, M., {Haugb{\o}lle}, T., \& {Nordlund}, {\r{A}}. 2017, \apj, 846, 7

\bibitem[{{Kuffmeier} {et~al.}(2023){Kuffmeier}, {Jensen}, \&
  {Haugb{\o}lle}}]{kuffmeier2023}
{Kuffmeier}, M., {Jensen}, S.~S., \& {Haugb{\o}lle}, T. 2023, European Physical
  Journal Plus, 138, 272

\bibitem[{{Kuznetsova} {et~al.}(2022){Kuznetsova}, {Bae}, {Hartmann}, \& {Mac
  Low}}]{kuznetsova2022}
{Kuznetsova}, A., {Bae}, J., {Hartmann}, L., \& {Mac Low}, M.-M. 2022, \apj,
  928, 92

\bibitem[{{Kwon} {et~al.}(2009){Kwon}, {Looney}, {Mundy}, {Chiang}, \&
  {Kemball}}]{kwon2009}
{Kwon}, W., {Looney}, L.~W., {Mundy}, L.~G., {Chiang}, H.-F., \& {Kemball},
  A.~J. 2009, \apj, 696, 841

\bibitem[{{Lam} {et~al.}(2019){Lam}, {Li}, {Chen}, {Tomida}, \&
  {Zhao}}]{lam2019}
{Lam}, K.~H., {Li}, Z.-Y., {Chen}, C.-Y., {Tomida}, K., \& {Zhao}, B. 2019,
  \mnras, 489, 5326

\bibitem[{{Larson}(1969)}]{larson1969}
{Larson}, R.~B. 1969, \mnras, 145, 271

\bibitem[{{Le Gouellec} {et~al.}(2019){Le Gouellec}, {Hull}, {Maury}, {Girart},
  {Tychoniec}, {Kristensen}, {Li}, {Louvet}, {Cortes}, \&
  {Rao}}]{legouellec2019}
{Le Gouellec}, V. J.~M., {Hull}, C. L.~H., {Maury}, A.~J., {et~al.} 2019, \apj,
  885, 106

\bibitem[{{Lebreuilly} {et~al.}(2021){Lebreuilly}, {Hennebelle}, {Colman},
  {Commer{\c{c}}on}, {Klessen}, {Maury}, {Molinari}, \&
  {Testi}}]{lebreuilly2021}
{Lebreuilly}, U., {Hennebelle}, P., {Colman}, T., {et~al.} 2021, \apjl, 917,
  L10

\bibitem[{{Lee} {et~al.}(2021){Lee}, {Charnoz}, \& {Hennebelle}}]{lee2021}
{Lee}, Y.-N., {Charnoz}, S., \& {Hennebelle}, P. 2021, \aap, 648, A101

\bibitem[{{Lesur} {et~al.}(2022){Lesur}, {Ercolano}, {Flock}, {Lin}, {Yang},
  {Barranco}, {Benitez-Llambay}, {Goodman}, {Johansen}, {Klahr}, {Laibe},
  {Lyra}, {Marcus}, {Nelson}, {Squire}, {Simon}, {Turner}, {Umurhan}, \&
  {Youdin}}]{lesur2022}
{Lesur}, G., {Ercolano}, B., {Flock}, M., {et~al.} 2022, arXiv e-prints,
  arXiv:2203.09821

\bibitem[{{Lesur} {et~al.}(2015){Lesur}, {Hennebelle}, \&
  {Fromang}}]{lesur2015}
{Lesur}, G., {Hennebelle}, P., \& {Fromang}, S. 2015, \aap, 582, L9

\bibitem[{{Li} {et~al.}(2013){Li}, {Krasnopolsky}, \& {Shang}}]{li2013}
{Li}, Z.-Y., {Krasnopolsky}, R., \& {Shang}, H. 2013, \apj, 774, 82

\bibitem[{{Manara} {et~al.}(2018){Manara}, {Morbidelli}, \&
  {Guillot}}]{manara2018}
{Manara}, C.~F., {Morbidelli}, A., \& {Guillot}, T. 2018, \aap, 618, L3

\bibitem[{{Masson} {et~al.}(2016){Masson}, {Chabrier}, {Hennebelle}, {Vaytet},
  \& {Commer{\c{c}}on}}]{masson2016}
{Masson}, J., {Chabrier}, G., {Hennebelle}, P., {Vaytet}, N., \&
  {Commer{\c{c}}on}, B. 2016, \aap, 587, A32

\bibitem[{{Matsumoto} {et~al.}(2017){Matsumoto}, {Machida}, \&
  {Inutsuka}}]{matsumoto2017}
{Matsumoto}, T., {Machida}, M.~N., \& {Inutsuka}, S.-i. 2017, \apj, 839, 69

\bibitem[{{Morbidelli} {et~al.}(2016){Morbidelli}, {Bitsch}, {Crida},
  {Gounelle}, {Guillot}, {Jacobson}, {Johansen}, {Lambrechts}, \&
  {Lega}}]{morbidelli2016}
{Morbidelli}, A., {Bitsch}, B., {Crida}, A., {et~al.} 2016, \icarus, 267, 368

\bibitem[{{Murillo} {et~al.}(2022){Murillo}, {van Dishoeck}, {Hacar},
  {Harsono}, \& {J{\o}rgensen}}]{murillo2022}
{Murillo}, N.~M., {van Dishoeck}, E.~F., {Hacar}, A., {Harsono}, D., \&
  {J{\o}rgensen}, J.~K. 2022, \aap, 658, A53

\bibitem[{{Najita} \& {Kenyon}(2014)}]{najita2014}
{Najita}, J.~R. \& {Kenyon}, S.~J. 2014, \mnras, 445, 3315

\bibitem[{{Pelkonen} {et~al.}(2021){Pelkonen}, {Padoan}, {Haugb{\o}lle}, \&
  {Nordlund}}]{pelkonen2021}
{Pelkonen}, V.~M., {Padoan}, P., {Haugb{\o}lle}, T., \& {Nordlund}, {\r{A}}.
  2021, \mnras, 504, 1219

\bibitem[{{Penston}(1969)}]{penston1969}
{Penston}, M.~V. 1969, \mnras, 144, 425

\bibitem[{{Pineda} {et~al.}(2023){Pineda}, {Arzoumanian}, {Andre}, {Friesen},
  {Zavagno}, {Clarke}, {Inoue}, {Chen}, {Lee}, {Soler}, \&
  {Kuffmeier}}]{pineda2023}
{Pineda}, J.~E., {Arzoumanian}, D., {Andre}, P., {et~al.} 2023, in Astronomical
  Society of the Pacific Conference Series, Vol. 534, Protostars and Planets
  VII, ed. S.~{Inutsuka}, Y.~{Aikawa}, T.~{Muto}, K.~{Tomida}, \& M.~{Tamura},
  233

\bibitem[{{Pineda} {et~al.}(2020){Pineda}, {Segura-Cox}, {Caselli},
  {Cunningham}, {Zhao}, {Schmiedeke}, {Maureira}, \& {Neri}}]{pineda2020}
{Pineda}, J.~E., {Segura-Cox}, D., {Caselli}, P., {et~al.} 2020, Nature
  Astronomy, 4, 1158

\bibitem[{{Pirogov} {et~al.}(2003){Pirogov}, {Zinchenko}, {Caselli},
  {Johansson}, \& {Myers}}]{pirogov2003}
{Pirogov}, L., {Zinchenko}, I., {Caselli}, P., {Johansson}, L.~E.~B., \&
  {Myers}, P.~C. 2003, \aap, 405, 639

\bibitem[{{Price} \& {Bate}(2007)}]{price2007}
{Price}, D.~J. \& {Bate}, M.~R. 2007, \mnras, 377, 77

\bibitem[{{Segura-Cox} {et~al.}(2020){Segura-Cox}, {Schmiedeke}, {Pineda},
  {Stephens}, {Fern{\'a}ndez-L{\'o}pez}, {Looney}, {Caselli}, {Li}, {Mundy},
  {Kwon}, \& {Harris}}]{seguracox2020}
{Segura-Cox}, D.~M., {Schmiedeke}, A., {Pineda}, J.~E., {et~al.} 2020, \nat,
  586, 228

\bibitem[{{Seifried} {et~al.}(2011){Seifried}, {Banerjee}, {Klessen}, {Duffin},
  \& {Pudritz}}]{seifried2011}
{Seifried}, D., {Banerjee}, R., {Klessen}, R.~S., {Duffin}, D., \& {Pudritz},
  R.~E. 2011, \mnras, 417, 1054

\bibitem[{{Seifried} {et~al.}(2012{\natexlab{a}}){Seifried}, {Banerjee},
  {Pudritz}, \& {Klessen}}]{seifried2012}
{Seifried}, D., {Banerjee}, R., {Pudritz}, R.~E., \& {Klessen}, R.~S.
  2012{\natexlab{a}}, \mnras, 423, L40

\bibitem[{{Seifried} {et~al.}(2013){Seifried}, {Banerjee}, {Pudritz}, \&
  {Klessen}}]{seifried2013}
{Seifried}, D., {Banerjee}, R., {Pudritz}, R.~E., \& {Klessen}, R.~S. 2013,
  \mnras, 432, 3320

\bibitem[{{Seifried} {et~al.}(2015){Seifried}, {Banerjee}, {Pudritz}, \&
  {Klessen}}]{seifried2015}
{Seifried}, D., {Banerjee}, R., {Pudritz}, R.~E., \& {Klessen}, R.~S. 2015,
  \mnras, 446, 2776

\bibitem[{{Seifried} {et~al.}(2012{\natexlab{b}}){Seifried}, {Pudritz},
  {Banerjee}, {Duffin}, \& {Klessen}}]{seifried2012out}
{Seifried}, D., {Pudritz}, R.~E., {Banerjee}, R., {Duffin}, D., \& {Klessen},
  R.~S. 2012{\natexlab{b}}, \mnras, 422, 347

\bibitem[{{Shu}(1977)}]{shu1977}
{Shu}, F.~H. 1977, \apj, 214, 488

\bibitem[{{Shu} {et~al.}(2006){Shu}, {Galli}, {Lizano}, \& {Cai}}]{shu2006}
{Shu}, F.~H., {Galli}, D., {Lizano}, S., \& {Cai}, M. 2006, \apj, 647, 382

\bibitem[{{Testi} {et~al.}(2014){Testi}, {Birnstiel}, {Ricci}, {Andrews},
  {Blum}, {Carpenter}, {Dominik}, {Isella}, {Natta}, {Williams}, \&
  {Wilner}}]{testi2014}
{Testi}, L., {Birnstiel}, T., {Ricci}, L., {et~al.} 2014, in Protostars and
  Planets VI, ed. H.~{Beuther}, R.~S. {Klessen}, C.~P. {Dullemond}, \&
  T.~{Henning}, 339--361

\bibitem[{{Teyssier}(2002)}]{teyssier2002}
{Teyssier}, R. 2002, \aap, 385, 337

\bibitem[{{Thieme} {et~al.}(2022){Thieme}, {Lai}, {Lin}, {Cheong}, {Lee},
  {Yen}, {Li}, {Lam}, \& {Zhao}}]{thieme2022}
{Thieme}, T.~J., {Lai}, S.-P., {Lin}, S.-J., {et~al.} 2022, \apj, 925, 32

\bibitem[{{Tobin} {et~al.}(2011){Tobin}, {Hartmann}, {Chiang}, {Looney},
  {Bergin}, {Chandler}, {Masqu{\'e}}, {Maret}, \& {Heitsch}}]{tobin2011}
{Tobin}, J.~J., {Hartmann}, L., {Chiang}, H.-F., {et~al.} 2011, \apj, 740, 45

\bibitem[{{Tobin} {et~al.}(2020){Tobin}, {Sheehan}, {Megeath},
  {D{\'\i}az-Rodr{\'\i}guez}, {Offner}, {Murillo}, {van 't Hoff}, {van
  Dishoeck}, {Osorio}, {Anglada}, {Furlan}, {Stutz}, {Reynolds}, {Karnath},
  {Fischer}, {Persson}, {Looney}, {Li}, {Stephens}, {Chandler}, {Cox},
  {Dunham}, {Tychoniec}, {Kama}, {Kratter}, {Kounkel}, {Mazur}, {Maud},
  {Patel}, {Perez}, {Sadavoy}, {Segura-Cox}, {Sharma}, {Stephenson}, {Watson},
  \& {Wyrowski}}]{tobin2020}
{Tobin}, J.~J., {Sheehan}, P.~D., {Megeath}, S.~T., {et~al.} 2020, \apj, 890,
  130

\bibitem[{Toro {et~al.}(1994)Toro, Spruce, \& Speares}]{toro1994}
Toro, E., Spruce, M., \& Speares, W. 1994, Shock Waves, 4, 25

\bibitem[{{Truelove} {et~al.}(1997){Truelove}, {Klein}, {McKee}, {Holliman},
  {Howell}, \& {Greenough}}]{truelove1997}
{Truelove}, J.~K., {Klein}, R.~I., {McKee}, C.~F., {et~al.} 1997, \apjl, 489,
  L179

\bibitem[{{Tsukamoto} {et~al.}(2015{\natexlab{a}}){Tsukamoto}, {Iwasaki},
  {Okuzumi}, {Machida}, \& {Inutsuka}}]{tsukamoto2015ohm}
{Tsukamoto}, Y., {Iwasaki}, K., {Okuzumi}, S., {Machida}, M.~N., \& {Inutsuka},
  S. 2015{\natexlab{a}}, \mnras, 452, 278

\bibitem[{{Tsukamoto} {et~al.}(2015{\natexlab{b}}){Tsukamoto}, {Takahashi},
  {Machida}, \& {Inutsuka}}]{tsukamoto2015rad}
{Tsukamoto}, Y., {Takahashi}, S.~Z., {Machida}, M.~N., \& {Inutsuka}, S.
  2015{\natexlab{b}}, \mnras, 446, 1175

\bibitem[{{Tychoniec} {et~al.}(2020){Tychoniec}, {Manara}, {Rosotti}, {van
  Dishoeck}, {Cridland}, {Hsieh}, {Murillo}, {Segura-Cox}, {van Terwisga}, \&
  {Tobin}}]{tychoniec2020}
{Tychoniec}, {\L}., {Manara}, C.~F., {Rosotti}, G.~P., {et~al.} 2020, \aap,
  640, A19

\bibitem[{{Tychoniec} {et~al.}(2018){Tychoniec}, {Tobin}, {Karska}, {Chandler},
  {Dunham}, {Harris}, {Kratter}, {Li}, {Looney}, {Melis}, {P{\'e}rez},
  {Sadavoy}, {Segura-Cox}, \& {van Dishoeck}}]{tychoniec2018}
{Tychoniec}, {\L}., {Tobin}, J.~J., {Karska}, A., {et~al.} 2018, \apjs, 238, 19

\bibitem[{{Ulrich}(1976)}]{ulrich1976}
{Ulrich}, R.~K. 1976, \apj, 210, 377

\bibitem[{{Valdivia-Mena} {et~al.}(2022){Valdivia-Mena}, {Pineda},
  {Segura-Cox}, {Caselli}, {Neri}, {L{\'o}pez-Sepulcre}, {Cunningham},
  {Bouscasse}, {Semenov}, {Henning}, {Pi{\'e}tu}, {Chapillon}, {Dutrey},
  {Fuente}, {Guilloteau}, {Hsieh}, {Jim{\'e}nez-Serra}, {Marino}, {Maureira},
  {Smirnov-Pinchukov}, {Tafalla}, \& {Zhao}}]{valdiviamena2022}
{Valdivia-Mena}, M.~T., {Pineda}, J.~E., {Segura-Cox}, D.~M., {et~al.} 2022,
  \aap, 667, A12

\bibitem[{{Valdivia-Mena} {et~al.}(2023){Valdivia-Mena}, {Pineda},
  {Segura-Cox}, {Caselli}, {Schmiedeke}, {Choudhury}, {Offner}, {Neri},
  {Goodman}, \& {Fuller}}]{valdiviamena2023}
{Valdivia-Mena}, M.~T., {Pineda}, J.~E., {Segura-Cox}, D.~M., {et~al.} 2023,
  \aap, 677, A92

\bibitem[{{Van Kooten} {et~al.}(2016){Van Kooten}, {Wielandt}, {Schiller},
  {Nagashima}, {Thomen}, {Larsen}, {Olsen}, {Nordlund}, {Krot}, \&
  {Bizzarro}}]{vankooten2016}
{Van Kooten}, E. M.~M.~E., {Wielandt}, D., {Schiller}, M., {et~al.} 2016,
  Proceedings of the National Academy of Science, 113, 2011

\bibitem[{{van Leer}(1977)}]{vanLeer1977}
{van Leer}, B. 1977, Journal of Computational Physics, 23, 276

\bibitem[{{van Leer}(1979)}]{vanLeer1979}
{van Leer}, B. 1979, Journal of Computational Physics, 32, 101

\bibitem[{{Walch} {et~al.}(2010){Walch}, {Naab}, {Whitworth}, {Burkert}, \&
  {Gritschneder}}]{walch2010}
{Walch}, S., {Naab}, T., {Whitworth}, A., {Burkert}, A., \& {Gritschneder}, M.
  2010, \mnras, 402, 2253

\bibitem[{{Wurster} \& {Bate}(2019)}]{wurster&bate2019}
{Wurster}, J. \& {Bate}, M.~R. 2019, \mnras, 486, 2587

\bibitem[{{Wurster} {et~al.}(2019){Wurster}, {Bate}, \& {Price}}]{wurster2019}
{Wurster}, J., {Bate}, M.~R., \& {Price}, D.~J. 2019, \mnras, 489, 1719

\bibitem[{{Yen} {et~al.}(2015){Yen}, {Koch}, {Takakuwa}, {Ho}, {Ohashi}, \&
  {Tang}}]{yen2015}
{Yen}, H.-W., {Koch}, P.~M., {Takakuwa}, S., {et~al.} 2015, \apj, 799, 193

\bibitem[{{Zhao} {et~al.}(2018){Zhao}, {Caselli}, {Li}, \&
  {Krasnopolsky}}]{zhao2018}
{Zhao}, B., {Caselli}, P., {Li}, Z.-Y., \& {Krasnopolsky}, R. 2018, \mnras,
  473, 4868

\bibitem[{{Zhao} {et~al.}(2016){Zhao}, {Caselli}, {Li}, {Krasnopolsky},
  {Shang}, \& {Nakamura}}]{zhao2016}
{Zhao}, B., {Caselli}, P., {Li}, Z.-Y., {et~al.} 2016, \mnras, 460, 2050

\end{thebibliography}

\end{document}